\documentclass[aps,preprint,prd,nofootinbib]{revtex4}
\pdfoutput=1
\usepackage{graphicx}
\usepackage{psfrag}

\usepackage{subfigure}
\usepackage{array}
\usepackage{graphicx}
\usepackage{amsmath}
\usepackage{amssymb}
\usepackage{mathrsfs}
\usepackage{bm}
\usepackage{slashed}
\usepackage{threeparttable}
\usepackage{multirow}
\usepackage[colorlinks, linkcolor=black, anchorcolor=black, citecolor=black]{hyperref}
\usepackage[amsmath,thmmarks,hyperref]{ntheorem}
\usepackage{soul}
\usepackage{ulem}  
\usepackage{upgreek}
\usepackage{amssymb}

\newcommand{\cancel}{\slashed}
\newcommand{\ud}{\mathrm{d}}
\newcommand{\dif}{\mathrm{d}}
\newcommand{\ui}{\mathrm{i}}
\newcommand{\tr}{\mathrm{tr}}
\newcommand{\Tr}{\mathrm{Tr}}
\newcommand{\e}{\mathrm{e}}
\newcommand{\C}{\mathcal C}
\newcommand{\D}{\mathcal D}
\newcommand{\la}{\mathcal{L}}

\allowdisplaybreaks[4]

\graphicspath{{figures/}} 

\begin{document}

\title{Calculation of the chiral Lagrangian coefficients with light vector mesons}
\author{Zi-Kan Geng$^1$\footnote{gzk19@mails.tsinghua.edu.cn}, Qing Wang$^{1,2}$\footnote{Corresponding author: wangq@mail.tsinghua.edu.cn}\\}
\address{$^1$Department of Physics, Tsinghua University, Beijing 100084, People's Republic of China\\
$^2$Center for High Energy Physics, Tsinghua University, Beijing 100084, People's Republic of China\\}

\baselineskip=20pt

\begin{abstract}

We calculate the coefficients in the effective chiral Lagrangian from QCD, which includes pseudo-scalar mesons and vector mesons (with hidden symmetry), up to $\mathcal O(p^4)$. This encompasses both the normal and anomalous parts. Our work builds on a previous study that derived the chiral Lagrangian from first principles of QCD, where the low-energy coefficients are defined in terms of specific Green's functions in QCD. This research extends our earlier efforts that focused on calculating the low-energy coefficients of the chiral Lagrangian for pure pseudo-scalar mesons. This marks the first calculation of chiral Lagrangian coefficients for vector mesons from QCD, particularly for the important parameters $a$ and $g$, which are typically considered inputs in existing literature. Notably, the regularization method used previously is inadequate for this broader scope. We find that cut-off regularization yields reasonable results for both pseudo-scalar mesons and vector mesons, though it has certain limitations. Finally, we demonstrate that our method aligns with the Weinberg sum rules.

\vspace*{0.5cm}


\end{abstract}

\maketitle

\section{Introduction}

The chiral Lagrangian for low lying pseudo-scalar mesons \cite{Weinberg:1978kz,Gasser:1983yg,Gasser:1984gg} stands as the most successful effective field theory, having been extended to $\mathcal O(p^8)$ \cite{Bijnens:2018lez} for higher precision in the low-energy expansion. To extend the energy reach of chiral perturbation theory (ChPT), more hadrons can be included as active degrees of freedom. Beyond the low-lying pseudo-scalar mesons, the next mesons in the hadron spectrum are the light vector mesons, the $\rho$ meson and its flavor partners. Several phenomenological models describe these vector mesons, including the matter field model \cite{Ecker:1989yg}, the massive Yang-Mills model \cite{Schwinger:1967tc,Kaymakcalan:1984bz,Meissner:1987ge}, the hidden local symmetry (HLS) model \cite{Bando:1987br,Harada:2003jx} and the anti-symmetric tensor field model \cite{Ecker:1988te}, among others. It has been shown that these models are equivalent at least at the tree level, and the relation of low energy coefficients (LECs) across different models have been studied \cite{Ecker:1989yg,Tanabashi:1995nz}. In the chiral Lagrangian, the number of the LECs increases rapidly with higher orders of the low-energy expansion or when more hadrons are included. For example in the three-flavor case, the $\mathcal O(p^2), \mathcal O(p^4), \mathcal O(p^6)$ and $\mathcal O(p^8)$ chiral Lagrangians for low lying pseudo-scalar mesons have 2, 12, 117 and 1959 LECs respectively \cite{Graf:2020yxt}. In the anti-symmetric tensor field model incorporating light vector mesons, the $\mathcal O(p^2), \mathcal O(p^4)$ and $\mathcal O(p^6)$ chiral Lagrangians for the three-flavor case have 4, 70 and 846 LECs respectively \cite{Guo:2020cmv}. Meanwhile, in the HLS model, the $\mathcal O(p^2), \mathcal O(p^4)$ and $\mathcal O(p^6)$ chiral Lagrangians for the three-flavor case have 4, 31 and 495 LECs respectively \cite{Guo:2020cmv}. Given the large number of LECs, it is challenging to fix them all from experimental data, making the calculation of LECs crucial for enhancing the predictive power of the chiral Lagrangian.

There are various methods to obtain LECs, with some of the more common ones being resonance chiral theory \cite{Ecker:1988te,Amoros:1999dp,Knecht:2001xc}, sum rules \cite{Golterman:2014nua}, lattice QCD \cite{Necco:2008ish,Aoki:2016frl}, holographic theory \cite{Harada:2006di,Harada:2010cn,Ma:2012zm}, QCD \cite{Wang:1999cp,Yang:2002hea,Jiang:2009uf,Jiang:2010wa} and global fit \cite{Bijnens:2011tb}. Some methods provide values for individual LECs, while others yield values for combinations of LECs. Each method has its own advantages and disadvantages. The most systematic and reliable approach involves studying the relationship between the chiral Lagrangian and the fundamental principles of QCD. In previous papers \cite{Wang:1999cp,Wang:2000mu}, we derived the effective chiral Lagrangian for pseudo-scalar mesons and vector mesons (with hidden symmetry) from the first principles of QCD without approximation. All the LECs are expressed in terms of specific Green's functions in QCD, which can be considered the fundamental QCD definitions of these coefficients. However, solving these expressions strictly is challenging. Therefore, we developed a systematic numerical method to calculate the LECs, which are expressed in terms of the quark self-energy $\Sigma(k^2)$ under certain approximations, such as the large-$N_c$ limit, the leading order in dynamical perturbation theory and the improved ladder approximation. By solving the Schwinger-Dyson equation (SDE) for $\Sigma(k^2)$, we obtain the approximate LECs for pseudo-scalar mesons up to the $\mathcal O(p^6)$ \cite{Yang:2002hea,Jiang:2009uf,Jiang:2010wa}, which are consistent with the experimental values. Compared to the similar studies which rely on certain a priori assumptions or approximations from the outset \cite{Balog:1984upv,Andrianov:1985ay,Klevansky:1992qe,Hatsuda:1994pi,Bijnens:1995ww,Holdom:1992fn}, the systematic errors in our calculations can be compensated step by step \cite{Jiang:2015dba}. 

Although our calculation for the LECs of pseudo-scalar mesons was successful, extending these calculations to vector mesons is nontrivial. The key difference lies in the introduction of HLS and the unique properties of vector mesons. Pseudo-scalar mesons are treated as Goldstone bosons following spontaneous chiral symmetry breaking. In the HLS model, vector mesons are considered gauge fields, with their masses arising from the Higgs mechanism. In previous papers \cite{Yang:2002hea,Lu:2002fx}, we used Schwinger proper time regularization to calculate the LECs for pseudo-scalar mesons, yielding results that matched those in the literature \cite{Espriu:1989ff} and appeared reasonable \cite{Yang:2002hea,Jiang:2009uf}. However, when this regularization was applied to the HLS model to calculate the LECs for vector mesons, the decay constant for vector mesons significantly deviated from experimental values. We explored the impact of different regularizations on the $\mathcal O(p^2)$ LECs and found that the cut-off regularization produced LECs for vector mesons consistent with experimental values, albeit at a cost. Notably, $\mathcal O(p^4)$ LECs are insensitive to the regularizations, and there was minimal change in the LECs for pseudo-scalar mesons. References \cite{Chen:2020jiq,Chen:2021qkx} also used cut-off regularization to calculate the LECs in the chiral Lagrangian for heavy-light mesons, obtaining the reasonable results. This suggests that cut-off regularization may be suitable for higher energy regions. In summary, the $\mathcal O(p^2)$ LECs are sensitive to regularizations, and Schwinger proper time regularization, previously used, is inadequate for vector mesons. While the new regularization method aligns with empirical data, it still presents some issues. Nonetheless, we demonstrate that our method is consistent with the Weinberg sum rules \cite{Weinberg:1967kj} at the end of this article.

It is worth noting that no new parameters are introduced for vector mesons in this paper, even though the number of LECs is larger than those for pure pseudo-scalar mesons. The parameters in the calculation are entirely fixed by pseudo-scalar mesons, as in Refs. \cite{Yang:2002hea,Jiang:2009uf}. For the most important coefficients $a$ and $g$, this is the first time that they have been obtained from theoretical calculation, whereas they are typically treated as inputs in the literature.

The paper is organized as follows. In Sec.\:\uppercase\expandafter{\romannumeral2}, we provide a brief review about the $\mathcal O(p^2)$ Lagrangian in HLS model. In Sec.\:\uppercase\expandafter{\romannumeral3}, we review
our previous derivation of the effective chiral Lagrangian with vector mesons from QCD. In Sec.\:\uppercase\expandafter{\romannumeral4}, we perform the momentum expansion and express the LECs in term of the quark's self-energy. In Sec.\:\uppercase\expandafter{\romannumeral5}, we present and discuss the numerical results. We derive the Weinberg sum rules using our method in Sec.\:\uppercase\expandafter{\romannumeral6}, and summarize in the end. The $\mathcal O(p^4)$ Lagrangian, including the normal and anomalous parts in HLS model, along with the analytical expressions for corresponding LECs, are included in the Appendix.

\section{The hidden local symmetry model of Vector Mesons}\label{sec:hlsmodel}

The hidden local symmetry model describing the vector mesons is based on the $G_{\text{global}}\times H_{\text{local}}$ symmetry, where $G=SU(3)_L\times SU(3)_R$ is the global chiral symmetry of the ChPT and $H=SU(3)_V$ is the HLS. The entire symmetry is spontaneously broken down to a diagonal sum $H$ which is the flavor symmetry. The gauge bosons of HLS acquire masses by the Higgs mechanism, and are then identified as the light vector mesons, $\rho$ meson and its flavor partners.

In the HLS model, the building block $U$ in the ChPT is split into two variables $\xi_{L}$ and $\xi_{R}$:
\begin{equation}
U=\xi_R^\dagger\xi_L.
\end{equation}
The ambiguity in this split  is known as the hidden local symmetry, $H_{local}$. Under the chiral symmetry and the HLS, these two variables transfrom as
\begin{equation}
\xi_{L,R}\to\xi'_{L,R}=h(x)\cdot\xi_{L,R}(x)\cdot g^\dagger_{L,R},
\end{equation}
where $h(x)\in [SU(3)_V]_{local},~g_{L,R}\in [SU(3)_L\times SU(3)_R]_{global}.$ These variables are parameterized as
\begin{equation}
\xi_{L,R}=\e^{\ui\sigma/F_\sigma}\e^{\pm \ui\pi/F_\pi},~~~\pi=\pi^aT_a,~\sigma=\sigma^aT_a,
\end{equation}
where $\pi$ denote the Goldstone bosons associated with the spontaneous chiral symmetry breaking and $\sigma$ denote the Goldstone bosons associated with the spontaneous breaking of HLS which are absorbed into the gauge bosons.

When gauging the chiral symmetry, we can construct two important building blocks that are chiral invariant and covariant under HLS:
\begin{equation}
\begin{aligned}
\hat\alpha_{\bot\mu}&=(D_\mu\xi_R\cdot\xi_R^\dagger-D_\mu\xi_L\cdot\xi_L^\dagger)/(2\ui),\\
\hat\alpha_{\parallel\mu}&=(D_\mu\xi_R\cdot\xi_R^\dagger+D_\mu\xi_L\cdot\xi_L^\dagger)/(2\ui),\\
\hat\alpha_{\bot\mu}&\to h(x)\hat\alpha_{\bot\mu}h^\dagger(x),~\hat\alpha_{\parallel\mu}\to h(x)\hat\alpha_{\parallel\mu}h^\dagger(x).
\end{aligned}
\end{equation}
The covariant derivatives are defined as
\begin{equation}
\begin{aligned}
D_\mu\xi_L&=\partial_\mu\xi_L-\ui V_\mu\xi_L+\ui\xi_L\la_\mu,\\
D_\mu\xi_R&=\partial_\mu\xi_R-\ui V_\mu\xi_R+\ui\xi_R\mathcal{R}_\mu,
\end{aligned}
\end{equation}
where $V_\mu=g\rho_\mu$ are the gauge fields of HLS, $g$ is the coupling of HLS, and $\la_\mu,\mathcal{R}_\mu$ are the external gauge fields of the chiral symmetry. These gauge fields transform as
\begin{equation}
\begin{aligned}
V_\mu&\to h(x)(V_\mu+\ui\partial_\mu)h^\dagger(x),\\
\la_\mu&\to g_L(\la_\mu+\ui\partial_\mu) g^\dagger_L,\\
\mathcal R_\mu&\to g_R(\mathcal R_\mu+\ui\partial_\mu)  g^\dagger_R.
\end{aligned}
\end{equation}
The order assignment of the new variables is
\begin{equation}
\xi_{L}\sim\xi_{R}\sim\mathcal O(1),~~~V_\mu\sim g\sim\mathcal O(p).
\end{equation}

The leading order Lagrangian is given by \cite{Harada:2003jx}
\begin{equation}
\begin{aligned}
\la&=\la_A+\la_\chi+a\la_V+\la_{\text{kin}}(V_\mu)\\
&=F^2_\pi \tr[\hat\alpha_{\bot\mu}\hat\alpha^\mu_\bot]+\frac{F_\pi^2}{4}\tr[\hat\chi+\hat\chi^\dagger]+F^2_\sigma \tr[\hat\alpha_{\parallel\mu}\hat\alpha^\mu_\parallel]-\frac{1}{2g^2}\tr[V_{\mu\nu}V^{\mu\nu}],\label{eq:p2hls}
\end{aligned}
\end{equation}
where $\hat\chi$ is the adjoint representation of the HLS for external source field $\chi$ using the "converters" $\xi_{L}$ and $\xi_{R}$ as $\hat\chi=\xi_R\chi\xi_L^\dagger=2B_0\xi_R(s+\ui p)\xi_L^\dagger$, and $V_{\mu\nu}$ is the gauge field strength of the HLS gauge boson defined by
\begin{equation}
V_{\mu\nu}=\partial_\mu V_\nu-\partial_\nu V_\mu-\ui[V_\mu,V_\nu].
\end{equation}

When the parameter $a\equiv F_\sigma^2/F_\pi^2$ is specifically chosen as 2, the above Lagrangian can reproduce several phenomenological facts, such as the universality of the $\rho$-coupling $g_{\rho\pi\pi}=g$, the $\rho$ meson dominance of the pion's electromagnetic form factor $g_{\gamma\pi\pi}=0$ and the KSRF formulas \cite{Kawarabayashi:1966kd,Riazuddin:1966sw}:
\begin{equation}
g_\rho=2F_\pi^2g_{\rho\pi\pi},~~~m_\rho^2=2g^2_{\rho\pi\pi}F_\pi^2.
\end{equation}
Similarly to the square of the pseudoscalar meson mass, HLS model assigns $\mathcal O(p^2)$ to the square of the vector meson mass: $m_\rho^2\sim \mathcal O(p^2)$, which is inconsistent with experimental values. However, it is emphasized in the Section 4.1 of Ref.~\cite{Harada:2003jx} that, ``thanks to the gauge invariance, the HLS makes possible the systematic expansion including the vector meson loops, particularly when the vector meson mass is light. It turns out that such a limit can actually be realized in QCD when the number of massless flavors $n_f$ becomes large as was demonstrated in Refs.~\cite{Harada:1999zj,Harada:2000kb}. Then one can perform the derivative expansion with including the vector mesons under such an extreme condition where the vector meson masses are small, and extrapolate the results to the real world $n_f=3$ where the vector meson masses take the experimental values.'' The detailed discussion on the power counting in the HLS model can be found in Ref. \cite{Harada:2003jx}. The $\mathcal O(p^4)$ Lagrangian in HLS model can be found in the Appendix \ref{sec:hlsp4l}.

\section{chiral effective Lagrangian including light vector mesons from QCD}

In this section, we review our previous derivation of the effective chiral
Lagrangian with vector mesons from QCD \cite{Wang:2000mu}. The start point of the derivation is the generating functional of QCD, with an external source $J(x)$ introduced for the composite light quark fields, given in Ref.\cite{Wang:1999cp}:
\begin{equation}
\begin{aligned}
Z[J]&=\int\mathcal D\bar\psi\mathcal D\psi \exp\left\{\ui\int d^4x[\bar\psi(i\cancel\partial+J)\psi]\right\}\int\D\Psi\D\bar\Psi\D A_\mu\Delta_F(A_\mu)\\
&\times\exp\left\{\ui\int d^4x\left[\la_{QCD}(A)-\frac{1}{2\xi}[F^i(A_\mu)]^2-g\mathcal I_i^\mu A_\mu^i+\bar\Psi(\ui\cancel\partial-M-g\cancel A)\Psi\right]\right\},\label{eq:start}
\end{aligned}
\end{equation}
The definitions and conventions for fields in Eq.~(\ref{eq:start}) are the same as those in Ref.\cite{Wang:1999cp}. After integrating out the quark and gluon fields, and integrating in the bilocal auxiliary field $\Phi^{\sigma\rho}(x,y)\sim(1/N_c)\bar\psi_\alpha^\sigma(x)\psi_\alpha^\rho(y)$, the result is 
\begin{equation}
Z[J]=\int\mathcal D\Phi\mathcal D\Pi \e^{\ui\Gamma_0[J,\Phi,\Pi]},
\end{equation}
where
\begin{equation}
\begin{aligned}
\Gamma_0[J,\Phi,\Pi]&= -\ui N_c\Tr\ln[\ui\cancel\partial+J-\Pi]+N_c\int d^4xd^4x'\Phi^{\sigma\rho}(x,x')\Pi^{\sigma\rho}(x,x')\\
&+N_c\sum_{n=2}^{\infty}\int d^4x_1\cdots d^4x_nd^4x'_1\cdots d^4x'_n\frac{(-\ui)^n(N_cg^2)^{n-1}}{n!}\\
&\times\bar G^{\sigma_1\cdots\sigma_n}_{\rho_1\cdots\rho_n}(x_1,x'_1,\cdots,x_n,x'_n)\Phi^{\sigma_1\rho_1}(x_1,x'_1)\cdots\Phi^{\sigma_n\rho_n}(x_n,x'_n).\label{eq:seff0}
\end{aligned}
\end{equation}
The detail of this derivation can be found in Ref.\cite{Wang:1999cp}. 

Further by integrating in the local pseudo-scalar mesons and vector meson fields, and integrating in the corresponding auxiliary field, the result  can be expressed as
\begin{equation}
Z=\int\D\xi_R\D\xi_L\D V\D\Xi\D\tilde V\D\Phi\D\Pi\delta(\xi_L^\dagger \xi_L-1)\delta(\xi_R^\dagger \xi_R-1)\delta(\det\xi_R-\det\xi_L)\e^{\ui\Gamma_2[\xi_R,\xi_L,V,J,\Xi,\tilde V,\Phi,\Pi]},
\end{equation}
where
\begin{equation}
\begin{aligned}
&\Gamma_2[\xi_R,\xi_L,V,J,\Xi,\tilde V,\Phi,\Pi]=\Gamma_0[J,\Phi,\Pi]+\ui\Gamma_I[\Phi]\\
+&\ui N_c\int d^4x\tr_f\left\{\Xi\left[\e^{-\ui\frac{\theta}{N_f}}\xi_R\tr_l(P_R\Phi^T)\xi_L^\dagger-\e^{\ui\frac{\theta}{N_f}}\xi_L\tr_l(P_L\Phi^T)\xi_R^\dagger\right]\right\}\\
+&\ui N_c\int d^4x\left\{V_\mu^{ab}+\left[(\xi_LP_R+\xi_RP_L)\left(\frac{\Pi}{4{\mu}^4}-\frac{\ui\cancel\partial}{4}\right)(\xi^\dagger_RP_R+\xi^\dagger_LP_L)\right]^{(a\xi)(b\zeta)}\gamma_\mu^{\zeta\xi}\right\}\tilde V^{\mu,ba}.\label{eq:gamma2}
\end{aligned}
\end{equation}
After formally finishing the integration over the auxiliary fields $\Phi,\Pi,\Xi$ and $\tilde V$, we obtain
\begin{equation}
Z=\int\D\xi_R\D\xi_L\D V\delta(\xi_L^\dagger \xi_L-1)\delta(\xi_R^\dagger \xi_R-1)\delta(\det\xi_R-\det\xi_L)\e^{\ui S_{\text{eff}}[\xi_R,\xi_L,V,J,\Xi_c,\tilde V_c,\Phi_c,\Pi_c]},
\end{equation}
where
\begin{equation}
\e^{\ui S_{\text{eff}}[\xi_R,\xi_L,V,J,\Xi_c,\tilde V_c,\Phi_c,\Pi_c]}=\int\D\Xi\D\tilde V\D\Phi\D\Pi\e^{\ui\Gamma_2[\xi_R,\xi_L,V,J,\Xi,\tilde V,\Phi,\Pi]}.
\end{equation}
The fields $\Xi_c,\tilde V_c,\Phi_c,\Pi_c$ are the corresponding classical field defined by
\begin{equation}
\begin{aligned}
\Phi_c&=\frac{\int\D\Xi\D\tilde V\D\Phi\D\Pi~\Phi~\e^{\ui\Gamma_2[\xi_R,\xi_L,V,J,\Xi,\tilde V,\Phi,\Pi]}}{\int\D\Xi\D\tilde V\D\Phi\D\Pi\e^{\ui\Gamma_2[\xi_R,\xi_L,V,J,\Xi,\tilde V,\Phi,\Pi]}},\\
\Pi_c&=\frac{\int\D\Xi\D\tilde V\D\Phi\D\Pi~\Pi~\e^{\ui\Gamma_2[\xi_R,\xi_L,V,J,\Xi,\tilde V,\Phi,\Pi]}}{\int\D\Xi\D\tilde V\D\Phi\D\Pi\e^{\ui\Gamma_2[\xi_R,\xi_L,V,J,\Xi,\tilde V,\Phi,\Pi]}},\\
\Xi_c&=\frac{\int\D\Xi\D\tilde V\D\Phi\D\Pi~\Xi~\e^{\ui\Gamma_2[\xi_R,\xi_L,V,J,\Xi,\tilde V,\Phi,\Pi]}}{\int\D\Xi\D\tilde V\D\Phi\D\Pi\e^{\ui\Gamma_2[\xi_R,\xi_L,V,J,\Xi,\tilde V,\Phi,\Pi]}},\\
\tilde V^\mu_c&=\frac{\int\D\Xi\D\tilde V\D\Phi\D\Pi~\tilde V^\mu~\e^{\ui\Gamma_2[\xi_R,\xi_L,V,J,\Xi,\tilde V,\Phi,\Pi]}}{\int\D\Xi\D\tilde V\D\Phi\D\Pi\e^{\ui\Gamma_2[\xi_R,\xi_L,V,J,\Xi,\tilde V,\Phi,\Pi]}},
\end{aligned}
\end{equation}
which satisfy the classical field equations:
\begin{equation}
\frac{\partial S_{\text{eff}}}{\partial\Phi^{\sigma\rho}_{c}}=\frac{\partial S_{\text{eff}}}{\partial\Pi^{\sigma\rho}_{c}}=\frac{\partial S_{\text{eff}}}{\partial\Xi^{\sigma\rho}_{c}}=\frac{\partial S_{\text{eff}}}{\partial\tilde V^{\mu}_{c}}=0.
\end{equation}

To simplify the result, we define the rotated source and fields denoted with subscript $\xi$:
\begin{equation}
\begin{aligned}
J_\xi(x)&\equiv[\xi_L(x)P_R+\xi_R(x)P_L][J(x)+\ui\cancel\partial][\xi^\dagger_R(x)P_R+\xi^\dagger_L(x)P_L],\\
\Phi^T_\xi(x,y)&\equiv[\xi_R(x)P_R+\xi_L(x)P_L]\Phi^T(x,y)[\xi^\dagger_L(y)P_R+\xi^\dagger_R(y)P_L],\\
\Pi^T_\xi(x,y)&\equiv[\xi_L(x)P_R+\xi_R(x)P_L]\Pi^T(x,y)[\xi^\dagger_R(y)P_R+\xi^\dagger_L(y)P_L],\\
V^\mu_\xi&\equiv V^\mu-\frac{1}{2}[\xi_R\ui\partial^\mu\xi_R^\dagger+\xi_L\ui\partial^\mu\xi^\dagger_L].
\end{aligned}
\end{equation}
After this rotation, there is no explicit $\xi_L,\xi_R$-dependence in Eq.~(\ref{eq:gamma2}), and all $\xi_L,\xi_R$-dependence is absorbed into variables with subscript $\xi$. The final expression for the generating functional becomes
\begin{equation}
\begin{aligned}
Z&=\int\D\xi_R\D\xi_L\D V\delta(\xi_L^\dagger \xi_L-1)\delta(\xi_R^\dagger \xi_R-1)\delta(\det\xi_R-\det\xi_L)\\
&\times\exp\{\ui S_{\text{eff}}[1,1,V_\xi,J_\xi,\Xi_c,\tilde V_c,\Phi_{\xi c},\Pi_{\xi c}]\},
\end{aligned}
\end{equation}
where
\begin{equation}
\e^{\ui S_{\text{eff}}[1,1,V_\xi,J_\xi,\Xi_c,\tilde V_c,\Phi_{\xi c},\Pi_{\xi c}]}=\int\D\Xi\D\tilde V\D\Phi_\xi\D\Pi_\xi\e^{\ui\Gamma_2[1,1,V_\xi,J_\xi,\Xi,\tilde V,\Phi_\xi,\Pi_\xi]},\label{eq:seff}
\end{equation}
and
\begin{equation}
\begin{aligned}
\Gamma_2[1,1,V_\xi,J_\xi,\Xi,\tilde V,\Phi_\xi,\Pi_\xi]=&\Gamma_0[J_\xi,\Phi_\xi,\Pi_\xi]+\ui\Gamma_I[\Phi_\xi]\\
+&\ui N_c\int d^4x\tr_f\left\{\Xi\left[\e^{-\ui\frac{\theta}{N_f}}\tr_l(P_R\Phi^T_\xi)-\e^{\ui\frac{\theta}{N_f}}\tr_l(P_L\Phi^T_\xi)\right]\right\}\\
+&\ui N_c\int d^4x\left[V_{\xi\mu}^{ab}+\frac{1}{4{\mu}^4}\Pi_\xi^{(a\xi)(b\zeta)}\gamma_\mu^{\zeta\xi}\right]\tilde V^{\mu,ba}.
\end{aligned}
\end{equation}
The measure of light quark field $\D\bar\psi\D\psi$ is not invariant under this rotation, and therefore the term $\Tr\ln[\ui\cancel\partial+\cancel J-\Pi]$ in $\Gamma_0$ becomes
\begin{equation}
\Tr\ln[\ui\cancel\partial+\cancel J-\Pi]=\Tr\ln[\ui\cancel\partial+\cancel J_\xi-\Pi_\xi]+\Gamma_{WZ}[\xi_R^\dagger\xi_L,l,r].\label{eq:anomaly}
\end{equation}
where $\Gamma_{WZ}[\xi_R^\dagger\xi_L,l,r]$ is the well-known Wess-Zumino term \cite{Harada:2003jx}.

The details of this derivation can be found in Ref.~\cite{Wang:2000mu}, where the definition of the vector mesons was given by
\begin{equation}
V_\mu^{ab}=-\left[(\xi_LP_R+\xi_RP_L)\left(\frac{\Pi}{4{\mu}^4}\right)(\xi^\dagger_RP_R+\xi^\dagger_LP_L)\right]^{(a\xi)(b\zeta)}\gamma_\mu^{\zeta\xi}.
\end{equation}
This transforms homogeneously under the HLS, indicating that it is not the gauge boson. In fact, the rotated vector external source $v_\xi$ acts as the HLS gauge boson because of the definition of the rotated external source $J_\xi$:
\begin{equation}
J_\xi(x)\equiv[\xi_L(x)P_R+\xi_R(x)P_L][J(x)+\ui\cancel\partial][\xi^\dagger_R(x)P_R+\xi^\dagger_L(x)P_L].
\end{equation}
In Ref.\cite{Wang:2000mu}, an attempt to compensate the above mismatch by adding an extra term in $V_\mu$ resulted in the non-equivalence between the derived chiral Lagrangian and those in the HLS model. Therefore, in this paper, the definition of the vector mesons is revised to
\begin{equation}
V_\mu^{ab}=-\left[(\xi_LP_R+\xi_RP_L)\left(\frac{\Pi}{4{\mu}^4}-\frac{\ui\cancel\partial}{4}\right)(\xi^\dagger_RP_R+\xi^\dagger_LP_L)\right]^{(a\xi)(b\zeta)}\gamma_\mu^{\zeta\xi},
\end{equation}
which transforms inhomogeneously under the HLS, confirming that it is indeed the corresponding gauge boson.

\section{the calculations of the $p^2$ order and $p^4$ order LECs}\label{sec:calculation}

Under the large-$N_c$ limit, the field integrations in Eq.(\ref{eq:seff}) can be carried out using the saddle point approximation with the classical field equations:
\begin{equation}
\begin{aligned}
&\frac{1}{4\mu^4}\delta^{(4)}(x-y)\widetilde{\cancel V}_c^{\sigma\rho}(x)+\Phi^{\sigma\rho}_{\xi c}(x,y)=-\ui[(\ui\cancel\partial+J_\xi-\Pi_{\xi c})^{-1}]^{\rho\sigma}(y,x),\\
&\widetilde\Xi^{\sigma\rho}(x)\delta^{(4)}(x-y)+\Pi^{\sigma\rho}_{\xi c}(x,y)+\sum_{n=1}^{\infty}\int d^4x_1\cdots d^4x_nd^4x'_1\cdots d^4x'_n\\
&\times\frac{(-\ui )^{n+1}(N_cg^2)^{n}}{n!}\bar G(x_1,x'_1,\cdots,x_n,x'_n)\Phi_{\xi c}(x_1,x'_1)\cdots\Phi_{\xi c}(x_n,x'_n)=0,\\
&\tr_l\left[\e^{-\ui\frac{\theta}{N_f}}P_R\Phi^T_{\xi c}-\e^{\ui\frac{\theta}{N_f}}P_L\Phi^T_{\xi c}\right]=0,\\
&\tr_l\left[V_{\xi,\mu}^{ab}(x)+\frac{1}{4{\mu}^4}\Pi_{\xi c}(x,x)^{(a\xi)(b\zeta)}(\gamma_\mu)^{\zeta\xi}\right]=0.\label{eq:fieldeq}
\end{aligned}
\end{equation}
These equations indicate that $\Phi_{\xi c}$ and $\Pi_{\xi c}$ serve as the quark propagator and the quark self-energy with the external sources $J_\xi$ respectively. Under the large-$N_c$ limit, the effective action is
\begin{equation}
\begin{aligned}
S_{\text{eff}}&=\Gamma_2[1,1,V_\xi,J_\xi,\Xi_c,\tilde V_c,\Phi_{\xi c},\Pi_{\xi c}]=\Gamma_0[J_\xi,\Phi_{\xi c},\Pi_{\xi c}]\\
&= -\ui N_c\Tr\ln[\ui\cancel\partial+J_\xi-\Pi_{\xi c}]+N_c\int d^4xd^4x'\Phi_{\xi c}^{\sigma\rho}(x,x')\Pi_{\xi c}^{\sigma\rho}(x,x')\\
&+N_c\sum_{n=2}^{\infty}\int d^4x_1\cdots d^4x_nd^4x'_1\cdots d^4x'_n\frac{(-\ui)^n(N_cg^2)^{n-1}}{n!}\\
&\times\bar G^{\sigma_1\cdots\sigma_n}_{\rho_1\cdots\rho_n}(x_1,x'_1,\cdots,x_n,x'_n)\Phi_{\xi c}^{\sigma_1\rho_1}(x_1,x'_1)\cdots\Phi_{\xi c}^{\sigma_n\rho_n}(x_n,x'_n).\label{eq:gamma0}
\end{aligned}
\end{equation}
The term $\Gamma_I$ is neglected because it is an $\mathcal O(1)$ term, and the last two terms in Eq.(\ref{eq:gamma2}) vanish due to Eq.(\ref{eq:fieldeq}). The second and third terms in Eq.(\ref{eq:gamma0}) are still too complex to calculate directly. Therefore, we consider only the leading order in dynamical perturbation, accounting for the QCD interaction in the SDE that lead to the nonperturbative solution of chiral symmetry breaking, while neglecting other QCD corrections in positive powers of $g$ (perturbative). This approximation yields the well-known Pagels-Stokar formula for the pion decay constant $f_\pi$ \cite{Pagels:1979hd}. At the leading order in dynamical perturbation, the third terms in Eq.(\ref{eq:gamma0}) can be neglected due to their dependence on positive powers of $g$. The second terms in Eq.(\ref{eq:gamma0}) can also be neglected because they are of the same order as the third term, as shown in Eq.(\ref{eq:fieldeq}). Thus, the effective action simplifies to
\begin{equation}
S_{\text{eff}}=-\ui N_c\Tr\ln[\ui\cancel\partial+J_\xi-\Pi_{\xi c}].
\end{equation}

We define a new external source $J'_\xi(x)$ and a auxiliary field $\Pi'_{\xi c}$ as
\begin{equation}
\begin{aligned}
J'_\xi(x)&\equiv J_\xi(x)-[\xi_L(x)P_R+\xi_R(x)P_L]\ui\cancel\partial[\xi^\dagger_R(x)P_R+\xi^\dagger_L(x)P_L],\\
\Pi'_{\xi c}&\equiv \Pi_{\xi c}+\cancel V-[\xi_L(x)P_R+\xi_R(x)P_L]\ui\cancel\partial[\xi^\dagger_R(x)P_R+\xi^\dagger_L(x)P_L].
\end{aligned}
\end{equation}
They both transform homogeneously under the HLS. After this rearrangement, the effective action becomes
\begin{equation}
S_{\text{eff}}=-\ui N_c\Tr\ln[\ui\cancel\partial+\cancel V+J'_\xi-\Pi'_{\xi c}].
\end{equation}
Eq.(\ref{eq:fieldeq}) indicates that $\Pi_{\xi c}$ functions as the  quark self-energy, and we maintain the identification from Ref.\cite{Yang:2002hea}:
\begin{equation}
{\Pi'}_{\xi c}^{\sigma\rho}(x,y)=[\Sigma(\bar\nabla_x^2)]^{\sigma\rho}\delta^4(x-y),
\end{equation}
where $\bar\nabla_x^\mu=\partial^\mu_x-\ui V^\mu$ is the covariant derivative ensuring the correct homogeneous transformation of ${\Pi'}_{\xi c}^{\sigma\rho}$ under the HLS. The effective action can then be written as
\begin{equation}
S_{\text{eff}}=-\ui N_c\Tr\ln[\ui\bar{\cancel\nabla}+J'_\xi-\Sigma(\bar\nabla_x^2)].
\end{equation} 

When the vector meson field vanishs, $\Sigma(-p^2)$ is the conventional quark self-energy obtained by solving the SDE. Retaining only the leading order in dynamical perturbation and applying the improved ladder approximation, the SDE in the Euclidean space-time is
\begin{equation}
\Sigma(p^2)-\frac{3N_c}{2}\int\frac{\ud^4p}{4\pi^3}\frac{\alpha_s[(p-q)]}{(p-q)^2}\frac{\Sigma(q^2)}{q^2+\Sigma^2(q^2)}=0.\label{eq:SDE}
\end{equation}


To calculate the LECs, we expand the effective action as a Taylor series:
\begin{align}
S_{\text{eff}}&=-\ui N_c\int \dif^4x\int\frac{\dif^4k}{(2\pi)^4}\tr\ln[\slashed k+\ui\bar{\slashed\nabla}+J'_\xi-\Sigma((-\ui k+\bar\nabla)^2)]\nonumber\\
&=-\ui N_c\int \dif^4x\int\frac{\dif^4k}{(2\pi)^4}\tr\ln[\slashed k-\Sigma(-k^2)+\ui\bar{\slashed\nabla}+J'_\xi-\Sigma_1(\bar\nabla)]\nonumber\\
&=-\ui N_c\int \dif^4x\int\frac{\dif^4k}{(2\pi)^4}\left\{\tr\ln[\slashed k-\Sigma]+\tr\ln[1+(\ui\bar{\slashed\nabla}+J'_\xi-\Sigma_1)(\slashed k+\Sigma)X]\right\}\nonumber\\
&\simeq-\ui N_c\int \dif^4x\int\frac{\dif^4k}{(2\pi)^4}\tr\left[\sum_{n=1}^{\infty}\frac{(-1)^{n+1}}{n}[(\ui\bar{\slashed\nabla}+J'_\xi-\Sigma_1)(\slashed k+\Sigma)X]^n\right],\label{eq:expanddier}
\end{align}
where `tr' includes the traces over spinor and flavour spacec, $J'_\xi=\slashed v'_\xi+\slashed a_\xi\gamma_5-s_\xi+\ui p_\xi\gamma_5$, $X=1/(k^2-\Sigma^2)$, $\Sigma=\Sigma(-k^2)$, and $\Sigma_1(\bar\nabla)\equiv\Sigma((-\ui k+\bar\nabla)^2)-\Sigma(-k^2)$.

After making the momentum expansion, the $\mathcal O(p^2)$ Lagrangian is 
\begin{equation}
\la_{(2)}=F^2_0 \tr[a_\xi^2]+F^2_0 B_0\tr[s_\xi]+F^2_\sigma \tr[{v'}_\xi^2].
\end{equation}
where $F_0$ is the pion decay constant in the chiral limit, and these coefficients are expressed as the functions of the quark self-energy in the Minkowski space-time:
\begin{equation}
\begin{aligned}
F_0^2B_0&=\ui N_c\int\frac{\dif ^4k}{(2\pi)^4}\frac{4\Sigma(-k^2)}{k^2-\Sigma^2(-k^2)},\\
F_0^2&=-\ui N_c\int\frac{\dif ^4k}{(2\pi)^4}\frac{k^2+2\Sigma^2(-k^2)}{[k^2-\Sigma^2(-k^2)]^2},\\
F_\sigma^2&=-\ui N_c\int\frac{\dif ^4k}{(2\pi)^4}\frac{k^2-2\Sigma^2(-k^2)}{[k^2-\Sigma^2(-k^2)]^2}.\label{eq:f02b0}
\end{aligned}
\end{equation} 
The expression for $F_0^2B_0$ in Eq.(\ref{eq:f02b0}) aligns with those in Refs. \cite{Aoki:1990eq,Holdom:1990iq,Roberts:1994dr}. The expression for $F_0^2$ is consistent with the well-known Pagels-Stokar formula, up to a total derivative term \cite{Wang:1999cp}.
Utilizing  the following identities,
\begin{equation}
a_{\xi}^{\mu}=\hat\alpha_{\bot}^\mu,~~{v'}_{\xi}^{\mu}=\hat\alpha_{\parallel}^{\mu},~~s_\xi=\frac{1}{4B_0}(\hat\chi+\hat\chi^\dagger),
\end{equation}
we can reproduce the leading order Lagrangian, excluding the kinetic energy term of the vector meson in Eq.(\ref{eq:p2hls}). The kinetic energy term of the vector meson belongs to $\mathcal O(p^4)$ in our calculation.

After making the momentum expansion, the $\mathcal O(p^4)$ Lagrangian is
\begin{equation}
\begin{aligned}
\la_{(4)}&=\tr_f[\C_2(\dif _\mu a^\mu)^2+\C_3(\dif ^\mu a^\nu-\dif ^\nu a^\mu)(\dif _\mu a_{\nu}-\dif _\nu a_{\mu})+\C_4a^4+\C_5a^\mu a^\nu a_{\mu}a_{\nu}\\
&+\C_6s^2+\C_7p^2+\C_8s a^2+\C_9V^{\mu\nu}V_{\mu\nu}-\ui\C_{10}V^{\mu\nu}a_{\mu}a_{\nu}+\C_{11}p\dif _\mu a^\mu\\
&+\C_{12}(\dif _\mu {v'}^\mu)^2+\C_{13}(\dif ^\mu {v'}^\nu-\dif ^\nu {v'}^\mu)(\dif _\mu {v'}_{\nu}-\dif _\nu {v'}_{\mu})+\C_{14}{v'}^4\\
&+\C_{15}{v'}^\mu {v'}^\nu {v'}_{\mu}{v'}_{\nu}+\C_{16}s {v'}^2-\ui\C_{17}V^{\mu\nu}{v'}_{\mu}{v'}_{\nu}+\C_{18}s\ud_\mu {v'}^\mu\\
&+\C_{19}a^2{v'}^2+\C_{20} a_{\mu}a_{\nu}{v'}^\mu {v'}^\nu+\C_{21} a_{\mu}a_{\nu}{v'}^\nu {v'}^\mu+\C_{22}[(a^\mu{v'}_{\mu})^2+({v'}^\mu a_{\mu})^2]\\
&+\C_{23}a_{\mu}{v'}_{\nu}a^\mu {v'}^\nu+\C_{24}[(a^\mu{v'}_{\mu})^2-({v'}^\mu a_{\mu})^2]+\C_{25}p( a^\mu{v'}_{\mu}-{v'}^\mu a_{\mu})\\
&+\C_{26}p( a^\mu{v'}_{\mu}+{v'}^\mu a_{\mu})+\C_{27}(\dif ^\mu {v'}^\nu-\dif ^\nu {v'}^\mu)a_{\mu} a_{\nu}+\C_{28}(\dif ^\mu {v'}^\nu-\dif ^\nu {v'}^\mu)V_{\mu\nu}\\
&+\C_{29}(\dif^\mu a^\nu-\dif^\nu a^\mu)(a_{\mu}{v'}_{\nu}+{v'}_{\mu}a_{\nu})+\C_{30}(\dif^\mu a^\nu-\dif^\nu a^\mu)(a_{\mu}{v'}_{\nu}-{v'}_{\mu}a_{\nu})\\
&+\C_{31}\dif_\mu a^\mu(a_{\nu}{v'}^\nu-{v'}^\nu a_{\nu})],\label{eq:lap4c}
\end{aligned}
\end{equation}
where the subscript $\xi$ of rotated external sources has been omitted for simplicity, and the coefficients $\C_i$ are expressed as the functions of the quark self-energy, detailed in the Appendix \ref{sec:resp4}. The corresponding items for $\C_2\sim\C_{11}$ are the terms for pure pseudo-scalar mesons, as in Refs.\cite{Yang:2002hea,Jiang:2009uf}. The corresponding items for $\C_{18},\C_{24},\C_{26},\C_{30}$ are not invariant under charge conjugation, and thus should not appear which is an oversight in Ref.\cite{Wang:2000mu}. As will be seen later, the results for these four coefficients are zero, confirming the accuracy of our calculation. Other terms account for vector mesons.
By applying the following identities:
\begin{equation}
\begin{aligned}
a_{\xi}^{\mu}&=\hat\alpha_{\bot}^\mu,~~{v'}_{\xi}^{\mu}=\hat\alpha_{\parallel}^{\mu},\\
s_\xi&=\frac{1}{4B_0}(\hat\chi+\hat\chi^\dagger),~~\ui p_\xi=\frac{1}{4B_0}(\hat\chi-\hat\chi^\dagger),\\
\dif^\mu {v'}^\nu_\xi-\dif^\nu {v'}^\mu_\xi&=\hat{\mathcal V}^{\mu\nu}-V^{\mu\nu}+\ui[{v'}_\xi^\mu,{v'}_\xi^\nu]+\ui[a_\xi^\mu,a_\xi^\nu]\\
\dif^\mu a^\nu_\xi-\dif^\nu a^\mu_\xi&=\hat{\mathcal A}^{\mu\nu}+\ui[{v'}_\xi^\mu,a_\xi^\nu]+\ui[a_\xi^\mu,{v'}_\xi^\nu],\label{eq:identities}
\end{aligned}
\end{equation}
and the equations of fields:
\begin{equation}
\begin{aligned}
\ud_\mu a_\xi^\mu&=-\ui(a-1)[v'_{\xi,\mu},a_\xi^\mu]+B_0\left(p_\xi-\frac{1}{N_f}\tr[p_\xi]\right)+\mathcal O(p^4),\\
\ud_\mu {v'}_\xi^\mu&=\mathcal O(p^4),
\end{aligned}
\end{equation}
we can reproduce the kinetic energy term of the vector meson and the next leading order Lagrangian in the HLS model:
\begin{equation}
\la_{(4)}=-\frac{1}{2g^2}\tr[V_{\mu\nu}V^{\mu\nu}]+\la_{(4)y}+\la_{(4)\omega}+\la_{(4)z},
\end{equation}
The relations between the coefficients $\C_i$ and the LECs of HLS model are given by
\begin{align}
-\frac{1}{2g^2}&=\C_9+\C_{13}-\C_{28},\nonumber\\
y_1&=\C_4+2\C_{13}-\ui\C_{27},~~y_2=\C_5-2\C_{13}+\ui\C_{27},~~y_3=2\C_{13}+\C_{14},\nonumber\\
y_4&=-2\C_{13}+\C_{15},~~y_5=4\C_3+\C_{19}-2\ui\C_{29},\nonumber\\
y_6&=4\C_3-2\C_{13}+\C_{20}+\ui\C_{27}+2\ui\C_{29},\nonumber\\
y_7&=2\C_{13}+[\C_{21}+2(a-1)^2\C_2-2\ui(a-1)\C_{31}]-\ui\C_{27},\nonumber\\
y_8&=-2\C_3+[\C_{22}-(a-1)^2\C_2+\ui(a-1)\C_{31}]-\ui\C_{29},\nonumber\\
y_9&=-4\C_3+\C_{23}+2\ui\C_{29},~~y_{10}=0,~~y_{11}=0,~~y_{12}=0,~~y_{13}=0,\nonumber\\
y_{14}&=0,~~y_{15}=0,~~y_{16}=0,~~y_{17}=0,~~y_{18}=0,\nonumber\\
\omega_1&=\frac{\C_8}{4B_0},~~\omega_2=0,~~\omega_3=\frac{\C_{16}}{4B_0},~~\omega_4=0,\nonumber\\
\omega_5&=\frac{\ui}{4B_0}\left[\C_{25}+\ui(a-1)B_0\C_2+\ui(a-1)\C_{11}+B_0\C_{31}\right],\nonumber\\
\omega_6&=\frac{\C_6}{16B_0^2},~~\omega_7=0,~~\omega_8=-\frac{1}{16B_0^2}\left[\C_7+B_0^2\C_2+B_0\C_{11}\right],\nonumber\\
\omega_9&=\frac{1}{16B_0^2}\left[\frac{1}{N_f}B_0^2\C_2+\frac{1}{N_f}B_0\C_{11}\right],\nonumber\\
z_1&=\C_{13},~~z_2=\C_3,~~z_3=-2\C_{13}+\C_{28},\nonumber\\
z_4&=-\C_{10}-4\C_{13}+\ui\C_{27}+2\C_{28},~~z_5=-\C_{17}-4\C_{13}+2\C_{28},\nonumber\\
z_6&=4\C_{13}-\ui\C_{27},~~z_7=4\C_{13},~~z_8=-4\C_3+\ui\C_{29}.
\end{align}

For the derivation of the Wess-Zumino anomalous term from QCD, refer to Refs.\cite{Ma:2003uv,Jiang:2010wa}. Besides the Wess-Zumino term, the $\mathcal O(p^4)$ anomalous Lagrangian can also emerge from the momentum expansion of the effective action:
\begin{equation}
\begin{aligned}
\la_{A,(4)}=&b_1\ui\epsilon^{\mu\nu\sigma\rho}\tr_f[a_\mu a_\nu a_\sigma v_\rho]+b_2\epsilon^{\mu\nu\sigma\rho}\tr_f[a_\mu v_\nu(\dif_\sigma {v'}_\rho-\dif_\rho {v'}_\sigma)\\
&+a_\mu(\dif_\sigma {v'}_\rho-\dif_\rho {v'}_\sigma)v_\nu]+b_3\epsilon^{\mu\nu\sigma\rho}\tr_f[a_\mu v_\nu V_{\sigma\rho}+a_\mu V_{\sigma\rho} v_\nu]\\
&+b_4\ui\epsilon^{\mu\nu\sigma\rho}\tr_f[a_\mu v_\nu v_\sigma v_\rho],\label{eq:laap4}
\end{aligned}
\end{equation}
where the coefficients $b_i$ are expressed as the functions of the quark self-energy in the Minkowski space-time:
\begin{equation}
\begin{aligned}
b_1&=-4\ui N_c\int\frac{\dif^4k}{(2\pi)^4}X^4(\Sigma^4+\Sigma^2k^2),\\
b_2&=\ui N_c\int\frac{\dif^4k}{(2\pi)^4}X^3\Sigma^2,\\
b_3&=2\ui N_c\int\frac{\dif^4k}{(2\pi)^4}X^3(\Sigma\Sigma'k^2+\Sigma^2),\\
b_4&=-4\ui N_c\int\frac{\dif^4k}{(2\pi)^4}X^3\Sigma^2,
\end{aligned}
\end{equation}
where $X=1/(k^2-\Sigma^2)$. Utilizing Eq.(\ref{eq:identities}) and the following identities:
\begin{equation}
\begin{aligned}
&\hat\alpha_L^\mu=v_\xi^\mu-a_\xi^\mu,~~\hat\alpha_R^\mu=v_\xi^\mu+a_\xi^\mu,~~\epsilon_{\mu\nu\sigma\rho}F_V^{\mu\nu}=\frac{1}{2}\epsilon_{\mu\nu\sigma\rho}V^{\mu\nu},\\
&\epsilon_{\mu\nu\sigma\rho}\hat F_L^{\mu\nu}=\frac{1}{2}\epsilon_{\mu\nu\sigma\rho}(\mathcal V^{\mu\nu}-\mathcal A^{\mu\nu}),~~\epsilon_{\mu\nu\sigma\rho}\hat F_R^{\mu\nu}=\frac{1}{2}\epsilon_{\mu\nu\sigma\rho}(\mathcal V^{\mu\nu}+\mathcal A^{\mu\nu}),
\end{aligned}
\end{equation}
we can reproduce the $\mathcal O(p^4)$ anomalous Lagrangian in Eq.(\ref{eq:laap4hls}) except for the Wess-Zumino term. The relations between the coefficients $b_i$ and the LECs $c_i$ in Eq.(\ref{eq:laap4hls}) are given by
\begin{equation}
\begin{aligned}
c_1=&-\frac{2\pi^2}{N_c}(b_1+4b_2+b_4),~~c_2=\frac{2\pi^2}{N_c}(b_1-b_4),\\
c_3=&-\frac{16\pi^2}{N_c}(b_3-b_2),~~c_4=-\frac{16\pi^2}{N_c}b_2.
\end{aligned}
\end{equation}

\section{numerical results and discussion}\label{sec:results}

\begin{figure}
  \centering
  \includegraphics[width=0.8\linewidth]{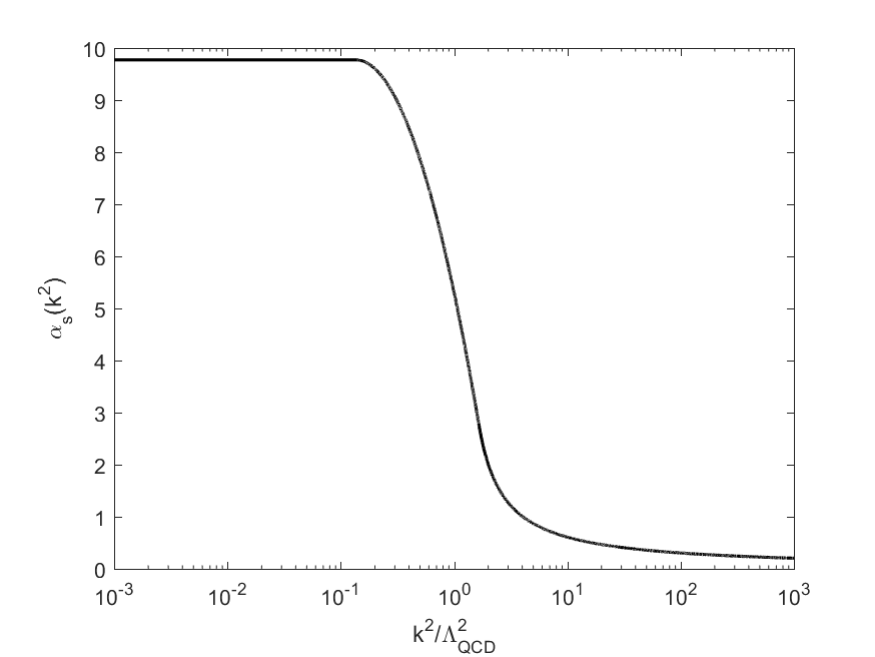}
  \caption{The running coupling constant $\alpha_s(p^2)$.}
  \label{fig:runningmodel}
\end{figure}

We choose the same running coupling constant $\alpha_s(p^2)$ as the ones in Refs.\cite{Yang:2002hea,Jiang:2009uf}:
\begin{displaymath}
\alpha_s(k^2)=\frac{12\pi}{33-2n_f}\times
\left\{ \begin{array}{cc}
7 & \ln(k^2/\Lambda^2_{\text{QCD}})\le-2\\
7-4[2+\ln(k^2/\Lambda^2_{\text{QCD}})]^2/5 & -2\le\ln(k^2/\Lambda^2_{\text{QCD}})\le0.5 \\
1/\ln(k^2/\Lambda^2_{\text{QCD}}) & \ln(k^2/\Lambda^2_{\text{QCD}})\ge0.5
\end{array}\right.
\end{displaymath}
which is plotted in FIG. \ref{fig:runningmodel}. It has the correct asymptotic behavior:
\begin{equation}
\alpha_s(k^2)\stackrel{k^2\to\infty}{\longrightarrow}\frac{12\pi}{33-2n_f}\frac{1}{\ln(k^2/\Lambda^2_{\text{QCD}})}.
\end{equation}
and there is only one parameter $\Lambda_{\text{QCD}}$.

\begin{figure}
  \centering
  \includegraphics[width=0.8\linewidth]{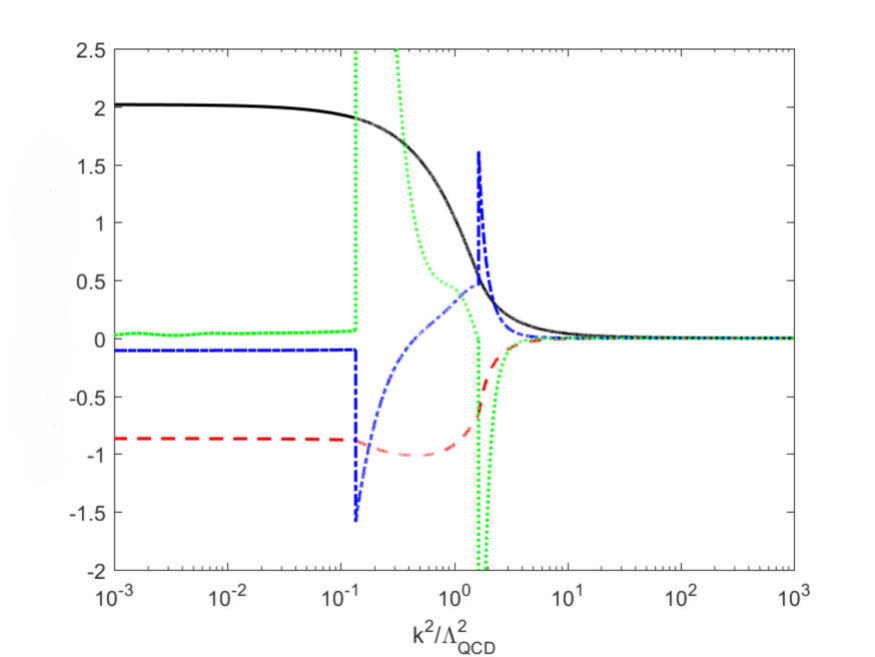}
  \caption{The obtained self-energy and its derivatives, where the solid line denotes the quark self-energy $\Sigma[k^2]/\Lambda_{\text{QCD}}$, the dashed line denotes the first derivative of quark self-energy $\Lambda_{\text{QCD}}\Sigma'[k^2]$, the dash-dotted line denotes the second derivative $\Lambda^3_{\text{QCD}}\Sigma''[k^2]$ and the dotted line denotes the third derivative $\Lambda^5_{\text{QCD}}\Sigma'''[k^2]$.}
  \label{fig:Sigma}
\end{figure}

With this running coupling constant, we solve the SDE for the quark self-energy, Eq.(\ref{eq:SDE}), numerically, and the details can be found in Ref.\cite{Yang:2002hea}. The obtained self-energy and its derivatives are plotted in FIG. \ref{fig:Sigma}, where the solid line represents the quark self-energy $\Sigma[k^2]/\Lambda_{\text{QCD}}$, the dashed line denotes the first derivative of quark self-energy $\Lambda_{\text{QCD}}\Sigma'[k^2]$, the dash-dotted line represents the second derivative $\Lambda^3_{\text{QCD}}\Sigma''[k^2]$ and the dotted line denotes the third derivative $\Lambda^5_{\text{QCD}}\Sigma'''[k^2]$. The higher-order derivatives of the quark self-energy are discontinuous, because the running coupling constant $\alpha_s(p^2)$ used in this paper is approximate. These discontinuities may introduce additional errors, so we retain only up to the second derivative of the self-energy and discard the higher order derivatives:
\begin{equation}
\Sigma_1(\bar\nabla)\simeq(-2\ui k\cdot\bar\nabla+\bar\nabla^2)\Sigma'(-k^2)+\frac{1}{2}(-2\ui k\cdot\bar\nabla+\bar\nabla^2)^2\Sigma''(-k^2).\label{eq:sigma1o2}
\end{equation}
This approximation will not affect the results for the $\mathcal O(p^2)$ coefficients.

By making use of the wick rotation $k^0\to\ui k^4$, the $p^2$ order LECs can be expressed as the functions of the quark self-energy in the Euclidean space-time:
\begin{equation}
\begin{aligned}
F_0^2B_0&=N_c\int\frac{\dif ^4k}{(2\pi)^4}\frac{4\Sigma(k^2)}{k^2+\Sigma^2(k^2)},\\
F_0^2&=-N_c\int\frac{\dif ^4k}{(2\pi)^4}\frac{k^2-2\Sigma^2(k^2)}{[k^2+\Sigma^2(k^2)]^2},\\
F_\sigma^2&=-N_c\int\frac{\dif ^4k}{(2\pi)^4}\frac{k^2+2\Sigma^2(k^2)}{[k^2+\Sigma^2(k^2)]^2}.\label{eq:f02b0ou}
\end{aligned}
\end{equation}
The asymptotic behavior of $\Sigma(k^2)$ reflecting chiral symmetry breaking is
\begin{equation}
\Sigma(k^2)\stackrel{k^2\to\infty}{\longrightarrow}\frac{\ln^{\gamma-1}(k^2/\Lambda^2_{\text{QCD}})}{k^2},\label{eq:sigmalargek}
\end{equation}
where $\gamma=9N_c/[2(33-2n_f)]$. With this asymptotic behavior of $\Sigma(k^2)$, it is evident that $F_0^2B_0$ exhibits logarithmic divergence, while both $F_0^2$ and $F_\sigma^2$ show quadratic divergence. Due to the lack of an analytical solution for the quark self-energy, the momentum integral in Eq.(\ref{eq:f02b0ou}) can not be performed analytically. This renders the commonly used regularizations in quantum field theory, such as dimensional regularization, inapplicable. In the literature, several regularizations are employed, including $\zeta$-function regularization \cite{Espriu:1989ff}, Schwinger proper time regularization \cite{Bijnens:1995ww,Lu:2002fx,Yang:2002hea,Jiang:2009uf} and cut-off regularization \cite{Chen:2020jiq,Chen:2021qkx}.
These regularizations involve a physical ultraviolet cut-off $\Lambda$ which should be on the order of the chiral symmetry breaking scale. If we assume $\Sigma(p^2)$ to be a constant mass $m$ as in Refs. \cite{Espriu:1989ff,Bijnens:1995ww}, the momentum integral in Eq.(\ref{eq:f02b0ou}) can be evaluated analytically, and the results under different regularizations are listed in TABLE \ref{tab:difregular}. 

The relation between the quark condensate $\langle\bar\psi\psi\rangle$ and $F_0^2B_0$ is given by
\begin{equation}
\langle\bar\psi\psi\rangle=-n_fF_0^2B_0.
\end{equation}
The chiral symmetry is spontaneously broken only if $F_0^2B_0$ is positive. In the $\zeta$-function regularization, which was the earliest regularization used in Ref. \cite{Espriu:1989ff}, $F_0^2B_0$ is logarithmically divergent but negative, making this regularization inappropriate. It was replaced with Schwinger proper time regularization in Ref. \cite{Bijnens:1995ww}. In the Schwinger proper time regularization, the logarithmic divergence in $F_0^2B_0$ is same as the ones in $\zeta$-function regularization, but the quadratic divergence in $F_0^2B_0$ is positive and dominant. As emphasized in the Chapter 6 of Ref.~\cite{Bijnens:1995ww}, ``The chiral symmetry is spontaneously broken by the quadratic divergence. In a regulator that does not have the quadratic divergence, like dimensional regularization, one always works in the phase where chiral symmetry is explicitly realized in the spectrum. In the ENJL model this means that we treat it as being in a phase with weakly interacting massive quarks.'' This discussion applies to QCD as well. The Schwinger proper time regularization, whether assuming a constant quark mass \cite{Bijnens:1995ww} or using a dynamical quark self-energy \cite{Yang:2002hea,Jiang:2009uf} has indeed provided reasonable LECs for pseudo-scalar mesons. 

However if the calculation for pseudo-scalar mesons in Ref. \cite{Yang:2002hea,Jiang:2009uf} is generalized to vector mesons, the obtained parameter $a\equiv F_\sigma^2/F_0^2\simeq0.35$. This result is significantly smaller than the phenomenological value $a_{\text{ph}}\sim2$, and violates many phenomenological facts discussed in Sec.\:\ref{sec:hlsmodel}. The Schwinger proper time regularization with a constant quark mass even yields a zero result for $F_\sigma^2$, as shown in TABLE \ref{tab:difregular}. Thus, Schwinger proper time regularization previously used does not work for vector mesons. In cut-off regularization, the divergence in $F_0^2B_0$ and the logarithmic divergence in $F_0^2$ are same as in the Schwinger proper time regularization, but the dominant contribution for $F_0^2$ and $F_\sigma^2$ becomes the quadratic divergence, which is absent in other regularizations. In this case, the obtained values of $F_0^2$ and $F_\sigma^2$ are reasonable, although their signs are negative.  As emphasized in the Section 4.5 of Ref.~\cite{Harada:2003jx}, ``In the usual phenomenological study in the ChPT of the pseudoscalar mesons as well as the calculations in the early stage of the ChPT with HLS, only the logarithmic divergence was included. As far as the bare theory is not referred to, the quadratic divergence is simply absorbed into redefinitions of the parameters. $\cdots$ However, as was shown in Refs.~\cite{Harada:1999zj,Harada:2000kb,Harada:2001rf}, the inclusion of the quadratic divergence is essential to studying the phase transition with referring to the bare theory. Moreover, it was shown \cite{Harada:2000at} that inclusion of the quadratic divergence is needed to match the HLS with the underlying QCD even for phenomenological reason." Quadratic divergence also appear in the bare parameters determined by Wilsonian matching of the EFT with the underlying QCD in Ref.~\cite{Harada:2000at}.  Specifically, the $\mathcal O(p^2)$ LECs are sensitive to the regularizations used. This suggests that the current handling of high-energy degrees of freedom in such research is still relatively crude. Most $\mathcal O(p^4)$ LECs, on the other hand, diverge logarithmically, and are therefore less sensitive to regularizations. 

\begin{table}
\centering
  \caption{The $\mathcal O(p^2)$ order coefficients under different regularizations when assuming $\Sigma(p^2)$ as a constant mass $m$}
  \begin{tabular}{cccc}
\hline\hline
 & $F_0^2B_0$ & $F_0^2$ & $F_\sigma^2$ \\
\hline
 cut-off regularization\cite{Chen:2020jiq,Chen:2021qkx} & $\frac{N_c}{4\pi^2}m\left[\Lambda^2-m^2\ln\frac{\Lambda^2}{m^2}\right]$ & $-\frac{N_c}{16\pi^2}\left[\Lambda^2-4m^2\ln\frac{\Lambda^2}{m^2}\right]$ & $-\frac{N_c}{16\pi^2}\Lambda^2$ \\
Schwinger proper time regularization\cite{Bijnens:1995ww,Lu:2002fx} & $\frac{N_c}{4\pi^2}m\left[\Lambda^2-m^2\ln\frac{\Lambda^2}{m^2}\right]$ & $\frac{N_c}{4\pi^2}m^2\ln\frac{\Lambda^2}{m^2}$ & 0 \\
$\zeta$-function regularization\cite{Espriu:1989ff} & $-\frac{N_c}{4\pi^2}m^3\ln\frac{\Lambda^2}{m^2}$ & $\frac{N_c}{4\pi^2}m^2\ln\frac{\Lambda^2}{m^2}$ &  \\
\hline\hline
\end{tabular}\label{tab:difregular}
\end{table}

In this paper, we adopt cut-off regularization, with the physical ultraviolet cutoff chosen as $\Lambda=1000^{+100}_{-100}$MeV. The pion decay constant in the chiral limit $|F_0^2|=(87\text{MeV})^2$ is used to determine the parameter $\Lambda_{\text{QCD}}$ in the running coupling constant, resulting in $\Lambda_{\text{QCD}}=537_{-94}^{+89}$MeV. The obtained quark condensate $\langle\bar\psi\psi\rangle$ is
\begin{equation}
\langle\bar\psi\psi\rangle=-n_fF_0^2B_0=-(336_{-51}^{+48}~\text{MeV})^3,
\end{equation}
which compares with the experimentally determined value $\langle\bar\psi\psi\rangle_\text{exp}=-(250~\text{MeV})^3$ from the QCD sum
rule at the typical hadronic mass scale \cite{Shifman:1978bx}. The important parameter $a$ in HLS model is calculated as
\begin{equation}
a\equiv\frac{F_\sigma^2}{F_0^2}=1.97_{-0.30}^{+0.34},\label{eq:resua}
\end{equation}
which is consistent with the phenomenological value $a_\text{ph}=2$. As discussed in Sec.\:\ref{sec:hlsmodel}, this result supports many phenomenological facts, for example the universality of the $\rho$-coupling $g_{\rho\pi\pi}=g$, the $\rho$ meson dominance of the pion's electromagnetic form factor $g_{\gamma\pi\pi}=0$ and the KSRF formulas. 

The obtained coefficients $\C_i$ in Eq.(\ref{eq:lap4c}) are listed in TABLE \ref{tab:cresults}. The value $\equiv0$ means that the corresponding item breaks charge conjugation invariance, indicating the correctness of our calculation. The coupling constant for HLS can be calculated as 
\begin{equation}
g=\left(-\frac{1}{2(\C_9+\C_{13}-\C_{28})}\right)^{\frac{1}{2}}=7.45_{-0.05}^{+0.08}.\label{eq:resug}
\end{equation}
This is the first calculation of important parameters $a$ and $g$ from the first principles of QCD, rather than treating them as input values as in the literature. Using these, we obtain the vector meson mass, the $\rho\pi\pi$ coupling and the $\rho-\gamma$ mixing strength \cite{Harada:2003jx} as
\begin{equation}
\begin{aligned}
M_V^2&=ag^2|F_0^2|=(910_{-79}^{+85}\text{MeV})^2,\\
g_{\rho\pi\pi}&=\frac{1}{2}ag=7.35_{-1.17}^{+1.34},\\
g_\rho&=ag|F^2_0|=0.111_{-0.018}^{+0.020}\text{GeV}^2.\label{eq:calresu}
\end{aligned}
\end{equation}
Considering the large theoretical uncertainty in this calculation, the obtained coupling for $\rho\pi\pi$ and the mixing strength for $\rho-\gamma$ are consistent with the experimental values \cite{Harada:2003jx},
\begin{equation}
g_{\rho\pi\pi}|_{\text{exp}}=6.00\pm0.01,~~g_\rho|_{\text{exp}}=0.119\pm0.001\text{GeV}^2.
\end{equation}
The obtained vector meson mass is slightly larger than the experimental value $m_\rho|_{\text{exp}}=775$MeV \cite{ParticleDataGroup:2022pth}, because the obtained coupling constant $g$ for HLS slightly larger than the phenomenological value $g\sim6$. We look forward to higher precision calculations like in Ref.\cite{Jiang:2015dba} to address this discrepancy.

\begin{table}
\centering
  \caption{The obtained coefficients $\C_i$ in Eq.(\ref{eq:lap4c}). $\C_6,\C_7$ are in units of $10^{-3}$GeV$^2$, $\C_8,\C_{11},\C_{16},\C_{25}$ are in units of $10^{-3}$GeV, the rest of coefficients are in units of $10^{-3}$. The resulting LECs are taken as the values at $\Lambda=1000$MeV with the superscript the difference caused at $\Lambda=1100$MeV and the subscript the difference caused at $\Lambda=900$MeV. The value $\equiv0$ means that the corresponding item breaks the invariance under charge conjugation.}
  \begin{tabular}{cccccc}
\hline\hline
$i$ & $\C_i$ & $j$ & $\C_j$ & $k$ & $\C_k$ \\
\hline
2 & $3.37_{+0.03}^{-0.04}$ & 12 & $10.0_{+0.6}^{-0.3}$ & 22 & $-15.2_{-2.2}^{+1.5}$ \\
3 & $-5.23_{-1.12}^{+0.76}$ & 13 & $-9.45_{-1.10}^{+0.74}$ & 23& $21.3_{+4.3}^{-2.8}$\\
4 & $-1.91_{-2.06}^{+1.30}$ & 14 & $-15.2_{-2.2}^{+1.5}$ & 24 & $\equiv0$ \\
5 & $7.35_{+2.13}^{-1.38}$ & 15 & $14.0_{+2.2}^{-1.5}$ & 25& $19.2_{-2.5}^{+2.3}\ui$ \\
6 & $18.8_{-1.2}^{+1.3}$ & 16 & $5.63_{-0.98}^{+0.93}$ & 26 & $\equiv0$ \\
7 & $26.2_{-3.5}^{+3.8}$ & 17 & $29.1_{+2.8}^{-2.0}$ & 27 & $9.69_{+4.31}^{-2.78}\ui$\\
8 & $21.5_{-2.1}^{+1.8}$ & 18 & $\equiv0$ & 28 & $-21.7_{-2.2}^{+1.6}$ \\
9 & $-21.2_{-1.3}^{+0.9}$ & 19 & $-20.8_{-4.4}^{+2.9}$  & 29 & $22.1_{+4.4}^{-2.9}\ui$ \\
10 & $-18.2_{-4.4}^{+2.9}$ & 20 & $33.1_{+8.6}^{-5.7}$ & 30 & $\equiv0$\\
11 & $32.0_{-4.2}^{+3.7}$ & 21 & $-7.48_{-4.26}^{+2.76}$ & 31  & $15.2_{+0.1}^{-0.1}\ui$\\
\hline\hline
\end{tabular}\label{tab:cresults}
\end{table}

The obtained values of $\mathcal O(p^4)$ LECs in HLS model are listed in TABLE \ref{tab:ywzresults}. Few studies provide these values. Ref.\cite{Ma:2012zm} calculated $\mathcal O(p^4)$ LECs in HLS model using two holographic QCD models: the Sakai-Sugimoto model and the BPS model. The resulting values are listed in TABLE \ref{tab:yzresultsma}. The outcomes from these two holographic QCD models show significant differences, and there is also a notable discrepancy compared to our results. We anticipate more experimental values or model calculations for the $\mathcal O(p^4)$ LECs of HLS model in the future.

In HLS model, the pion electromagnetic form factor $F_V^{\pi^\pm}$ is constructed from two contributions, one from $\gamma\pi\pi$ direct coupling diagram, another from $\rho$-mediated diagram:
\begin{equation}
F_V^{\pi^\pm}(Q^2)=g_{\gamma\pi\pi}(Q^2)+\frac{g_\rho(Q^2)g_{\rho\pi\pi}(Q^2)}{m_\rho^2+Q^2}.
\end{equation}
By using the Lagrangian given in Eq.~(\ref{eq:p2hls}) and Eq.~(\ref{eq:lp4z}), the above expression can be rewritten as \cite{Harada:2010cn}
\begin{equation}
F_V^{\pi^\pm}(Q^2)=\left(1-\frac{\tilde a}{2}\right)+\tilde z\frac{Q^2}{m_\rho^2}+\frac{\tilde a}{2}\frac{m_\rho^2}{m_\rho^2+Q^2},
\end{equation}
where
\begin{equation}
\begin{aligned}
\tilde a&=a\left(1-\frac{g^2z_4}{2}-g^2z_3+\frac{(g^2z_3)(g^2z_4)}{2}\right),\\
\tilde z&=\frac{1}{4}a(g^2z_6+(g^2z_3)(g^2z_4)).
\end{aligned}
\end{equation}
Using the results provided in Eq.~(\ref{eq:resua}), Eq.~(\ref{eq:resug}) and 
TABLE \ref{tab:ywzresults}, the obtained values are $\tilde a=2.09_{-0.31}^{+0.35},~~\tilde z=-0.781_{+0.123}^{-0.148}$. In Ref.\cite{Harada:2010cn}, $\tilde a$ and $\tilde z$ are determined in the holographic QCD model (Sakai-Sugimoto model) as $\tilde a_{\text{SS}}\simeq2.62,~\tilde z_{\text{SS}}\simeq0.08$. The values fitting the experimental data are $\tilde a_{\text{fit}}=2.44,~\tilde z_{\text{fit}}=0.08$. By comparison, our calculated $\tilde a$ is consistent with other results within the error range, but our $\tilde z$ shows some deviation from other results.

The relations between the coefficients in the HLS model and the anti-symmetric tensor field model are given by \cite{Tanabashi:1995nz}
\begin{equation}
a=\frac{F_V^2}{F_0^2},~~g=\frac{M_V}{F_V},~~z_4=\frac{2F_V(F_V-2G_V)}{M_V^2}.
\end{equation}
With these relations, we obtain the coefficients $F_V$ and $G_V$ in the anti-symmetric tensor field model:
\begin{equation}
F_V=122^{+10}_{-10}\text{MeV},~~G_V=56.0^{+4.5}_{-4.3}\text{MeV}.
\end{equation}
Considering the pion decay constant in this calculation is taken in the chiral limit, these coefficients are consistent with the experimental values $F_V|_\text{exp}=154$MeV, $G_V|_\text{exp}=69$MeV \cite{Ecker:1988te}.

\begin{table}
\centering
  \caption{The obtained values of $\mathcal O(p^4)$ LECs. The LECs are in units of $10^{-3}$. The resulting LECs are taken as the values at $\Lambda=1000$MeV with the superscript the difference caused at $\Lambda=1100$MeV and the subscript the difference caused at $\Lambda=900$MeV. The value $\equiv0$ means that the constants vanish at the large-$N_c$ limit.}
  \begin{tabular}{cccc}
\hline\hline
 $i$ & $y_i$ & $\omega_i$ & $z_i$  \\
\hline
1 & $-11.1_{-0.1}^{+0.1}$ & $3.22_{+0.29}^{-0.23}$ & $-9.44_{-1.11}^{+0.73}$ \\ 
2 & $16.5_{+0.1}^{-0.2}$ & $\equiv0$ & $-5.23_{-1.12}^{+0.76}$\\
3 & $-34.1_{-4.4}^{+2.9}$ & $0.844_{+0.001}^{-0.002}$ & $-2.77_{-0.05}^{+0.06}$ \\
4 & $32.9_{+4.4}^{-2.9}$ & $\equiv0$ & $2.97_{-0.02}^{+0.01}$  \\
5 & $2.57_{-0.10}^{+0.10}$ & $-12.2_{-0.4}^{+0.4}$ & $-34.6_{-3.0}^{+2.0}$\\
6 & $-22.9_{-6.6}^{+4.4}$ & $0.422_{+0.159}^{-0.091}$ & $-28.1_{-0.2}^{+0.2}$\\
7 & $19.2_{-2.1}^{+1.3}$ & $\equiv0$ & $-37.8_{-4.4}^{+2.9}$\\
8 & $-0.567_{+4.37}^{-2.88}$ & $0.401_{-0.101}^{+0.043}$ & $-1.21_{+0.09}^{-0.11}$\\
9 & $-1.98_{+0.02}^{-0.04}$ & $-0.329_{-0.020}^{+0.016}$  & \\
10$\sim$18 & $\equiv0$ &  & \\ 
\hline\hline
\end{tabular}\label{tab:ywzresults}
\end{table}

\begin{table}
\centering
  \caption{The obtained values of $\mathcal O(p^4)$ LECs from two holographic QCD models\cite{Ma:2012zm}. The LECs are in units of $10^{-3}$.}
  \begin{tabular}{ccccc}
\hline\hline
 $i$ & $y_i$(SS) & $y_i$(BPS) & $z_i$(SS) & $z_i$(BPS)\\
\hline
1 & $-1.10$ & $-71.9$ & & \\ 
2 & $1.10$ & $71.9$ & &\\
3 & $-2.83$ & $-154$ & & \\
4 & $2.83$ & $154$ &  $10.8$ & $90.3$\\
5 & $-15.9$ & $-12.3$ & $-7.33$ & $-131$\\
6 & $13.7$ & $-197$ & &\\
7 & $2.21$ & $209$ & &\\
8 & $-7.96$ & $-6.14$ & &\\
9 & $15.9$ & $12.3$  & &\\
\hline\hline
\end{tabular}\label{tab:yzresultsma}
\end{table}

Using of Eq.(\ref{eq:wyz2L}), we obtain the contributions of vector mesons to the $\mathcal O(p^4)$ Lagrangian for pure pseudo-scalar mesons. The obtained values of $L_i$ and their comparison with values in the literature are listed in TABLE \ref{tab:Lresults}. Our results are consistent with those in the literature.

\begin{table}
\centering
  \caption{The values of $L_i$ obtained by Eq.(\ref{eq:wyz2L}) and the values in the literature, which are in units of $10^{-3}$. The value $\equiv0$ means that the constants vanish at the large-$N_c$ limit.}
  \begin{tabular}{cccccc}
\hline\hline
 & This paper  & Ref.\cite{Jiang:2009uf} & Ref.\cite{Gasser:1984gg} & Ref.\cite{Pich:2008xj} & Ref.\cite{Bijnens:2011tb}\\
\hline
$L_1$ & $0.987_{+0.011}^{-0.013}$ & $1.23^{+0.03}_{-0.04}$  & 0.9$\pm$0.3 & 0.4$\pm$0.3 & 0.88$\pm$0.09\\
$L_2$ & $1.97_{+0.03}^{-0.02}$ & $2.46^{+0.05}_{-0.08}$ & 1.7$\pm$0.7 & 1.4$\pm$0.3 & 0.61$\pm$0.20\\
$L_3$ & $-5.58_{-0.06}^{+0.07}$ & $-6.85^{-0.14}_{+0.21}$  & -4.4$\pm$2.5 & -3.5$\pm$1.1 & -3.04$\pm$0.43\\
$L_4$ & $\equiv0$ & $\equiv0$ & 0$\pm$0.5 & -0.3$\pm$0.5 & 0.75$\pm$0.75\\
$L_5$ & $0.806_{+0.072}^{-0.060}$ & $1.48^{-0.01}_{-0.03}$  & 2.2$\pm$0.5 & 1.4$\pm$0.5 & 0.58$\pm$0.13\\
$L_6$ & $\equiv0$ & $\equiv0$ & 0$\pm$0.3 & -0.2$\pm$0.3 &0.29$\pm$0.85\\
$L_7$ & $-0.329_{-0.020}^{+0.017}$ & $-0.51^{+0.05}_{-0.06}$ & -0.4$\pm$0.15 & -0.4$\pm$0.2 &-0.11$\pm$0.15\\
$L_8$ & $0.823_{+0.058}^{-0.048}$ & $1.02^{-0.06}_{+0.06}$ & 1.1$\pm$0.3 & 0.9$\pm$0.3 &0.18$\pm$0.18\\
$L_9$ & $8.33_{+0.10}^{-0.12}$ & $8.86^{+0.24}_{-0.37}$ & 7.4$\pm$0.7 & 6.9$\pm$0.7 &5.93$\pm$4.30\\
$L_{10}$ & $-8.00_{-0.08}^{+0.11}$ & $-7.40^{-0.29}_{+0.44}$ & -6.0$\pm$0.7 & -5.5$\pm$0.7 & -4.06$\pm$0.39\\
\hline\hline
\end{tabular}\label{tab:Lresults}
\end{table}

The obtained values of $\mathcal O(p^4)$ anomalous LECs in HLS model and their comparison with values in the literature are listed in TABLE \ref{tab:anomcoe}. The results from the two holographic QCD models differ significantly, and there is also a significant difference compared to our results. By using the Lagrangian given in Eq.~(\ref{eq:p2hls}), Eq.~(\ref{eq:lp4z}) and Eq.~(\ref{eq:laap4hls}), effective $\omega\pi^0\gamma$ vertex function can be written as \cite{Harada:2010cn}
\begin{equation}
\Gamma^{\mu\nu}[\omega_\mu(p),\pi^0,\gamma^*_\nu(k)]=\frac{eN_c}{8\pi^2F_\pi}\epsilon^{\mu\nu\alpha\beta}p_\alpha k_\beta\cdot[A^{\omega\pi\gamma}+B^{\omega\pi\gamma}D_\rho(k^2)],
\end{equation}
where
\begin{equation}
\begin{aligned}
A^{\omega\pi\gamma}=&\frac{1}{2}g[(c_4+c_3\cdot g^2z_3)-c_3(1-g^2z_3)],\\
B^{\omega\pi\gamma}=&gc_3(1-g^2z_3),~~D_\rho(k^2)=\frac{m^2_\rho}{m^2_\rho-p^2}.
\end{aligned}
\end{equation}
The $\omega-\pi^0$ transition form factor $F_{\omega\pi^0}$ can be extracted from the $\omega\to\pi^0l^+l^- (l^\pm=e^\pm,\mu^\pm)$ decay width \cite{Harada:2010cn}:
\begin{equation}
F_{\omega\pi^0}(q^2)=(1-\tilde r)+\tilde rD_\rho(q^2),
\end{equation}
where $\tilde r=B^{\omega\pi\gamma}/(A^{\omega\pi\gamma}+B^{\omega\pi\gamma})$. Using the results given in Eq.~(\ref{eq:resug}), TABLE \ref{tab:ywzresults} and TABLE \ref{tab:anomcoe}, the obtained value is $\tilde r=1.77$. In Ref.\cite{Harada:2010cn}, $\tilde r$ is determined in the holographic QCD model (Sakai-Sugimoto model) as $\tilde r_{\text{SS}}\simeq1.53$, and the value by fitting to the experimental data is $\tilde r_{\text{fit}}=2.08$. By comparison, our calculated $\tilde r$ is consistent with other results.

\begin{table}
\centering
  \caption{The obtained values of $\mathcal O(p^4)$ anomalous LECs and the values in the literature.}
  \begin{tabular}{ccccc}
\hline\hline
 & $c_1$ & $c_2$ & $c_3$ & $c_4$\\
\hline
This paper & 0.0389 & 0.190 & -1.50 & -0.458 \\
SS model\cite{Ma:2012zm} & 0.382 & -0.130 & 0.767 & --\\
BPS model\cite{Ma:2012zm} & -0.207 & 3.03 & 1.47 & --\\
\hline\hline
\end{tabular}\label{tab:anomcoe}
\end{table}

\section{Weinberg sum rules}

Although the signs of obtained $F_0^2$ and $F_\sigma^2$ are negative, our method is consistent with the Weinberg sum rules \cite{Weinberg:1967kj} which relate the spectral functions of the vector and axial-vector currents. These spectral functions, $\rho_V(p^2)$ and $\rho_A(p^2)$, are defined by the quark current Green functions:
\begin{equation}
\begin{aligned}
\langle V_i^\mu(x)V_j^\nu(0)\rangle&=(2\pi)^{-3}\delta_{ij}\int\ud^4p\theta(p^0)\e^{-\ui p\cdot x}\rho_V(p^2)[-g^{\mu\nu}+p^\mu p^\nu/p^2],\\
\langle A_i^\mu(x)A_j^\nu(0)\rangle&=(2\pi)^{-3}\delta_{ij}\int\ud^4p\theta(p^0)\e^{-\ui p\cdot x}\{\rho_A(p^2)[-g^{\mu\nu}+p^\mu p^\nu/p^2]+F_\pi^2\delta(p^2)p^\mu p^\nu\},\label{eq:currentgf}
\end{aligned}
\end{equation}
where $V_i^\mu(x)$ and $A_i^\mu(x)$ are vector and axial-vector currents respectively. The first and second Weinberg sum rules are given by
\begin{equation}
\begin{aligned}
&\int_{0}^\infty[\rho_V(\mu^2)-\rho_A(\mu^2)]\mu^{-2}d\mu^{2}=F_\pi^2,\\
&\int_{0}^\infty[\rho_V(\mu^2)-\rho_A(\mu^2)]d\mu^{2}=0,
\end{aligned}
\end{equation}
respectively, where $p^2=\mu^2$. 

In this section, we will derive the Weinberg sum rules using our method. We start with Eq.(\ref{eq:seff0}), which can be regarded as the deformed action of QCD after integrating out the quark and gluon fields, and integrating in the bilocal auxiliary fields $\Phi^{\sigma\rho}(x,y)$ and $\Pi^{\sigma\rho}(x,y)$. In this case, there is no need to integrate in the local pseudo-scalar mesons and vector meson fields. After formally completing the integration over the auxiliary fields $\Phi,\Pi$, taking the large-$N_c$ limit and the leading order in dynamical perturbation, the generating functional of quark current Green functions simplifies to
\begin{equation}
W[J]=-\ui N_c\Tr\ln[\ui\cancel\partial+J-\Pi_{c}],
\end{equation} 
where $J(x)=\slashed v(x)+\slashed a(x)\gamma^5-s(x)+\ui p(x)\gamma^5$ is the external source for quark current, and $\Pi_c$ is the classical field of auxiliary field $\Pi^{\sigma\rho}(x,y)$ which satisfies the classical field equation:
\begin{equation}
\Phi^{\sigma\rho}_{c}(x,y)=-\ui[(\ui\cancel\partial+J-\Pi_{c})^{-1}]^{\rho\sigma}(y,x).
\end{equation}

Firstly, we calculte the vector current Green function:
\begin{equation}
\begin{aligned}
\langle V_i^\mu(x)V_j^\nu(y)\rangle&=\left.\frac{\delta^2 W}{\delta v^i_\mu(x)\delta v^j_\nu(y)}\right|_{J=0}\\
&=-\ui N_c\ud^4z\ud^4z'\tr_{lf}\left.\left[\gamma^\mu\lambda^i\Phi_c(x,z)\frac{\delta\Phi_c^{-1}(z,z')}{\delta v^j_\nu(y)}\Phi_c(z',x)\right]\right|_{J=0}\\
&=-\ui N_c\ud^4z\ud^4z'\tr_{lf}\left[\gamma^\mu\lambda^i\Phi_c(x,z)\delta(z-z')\delta(z-y)\gamma^\nu\lambda^j\Phi_c(z',x)\right]|_{J=0}\\
&=-\ui N_c\tr_{lf}\left[\gamma^\mu\lambda^i\Phi_c(x,y)\gamma^\nu\lambda^j\Phi_c(y,x)\right]|_{J=0},
\end{aligned}
\end{equation}
where $\lambda^i$ is the Gell-Mann matrix in flavor space.
As in section \ref{sec:calculation}, $\Pi_c$ can be further approximated as the quark self-energy:
\begin{equation}
\Pi_c(x,y)=\int\frac{\ud^4p}{(2\pi)^4}\e^{-\ui p\cdot(x-y)}\Sigma(-p^2).
\end{equation}
Then the vector current Green function is
\begin{equation}
\begin{aligned}
&\langle V_i^\mu(x)V_j^\nu(y)\rangle\\
&=-\ui N_c\delta^{ij}\int\frac{\ud^4p\ud^4q}{(2\pi)^8}\e^{-\ui(p-q)\cdot(x-y)}\tr_{l}\left[\gamma^\mu\frac{1}{\slashed p-\Sigma(-p^2)}\gamma^\nu\frac{1}{\slashed q-\Sigma(-q^2)}\right]\\
&=-\ui N_c\delta^{ij}\int\frac{\ud^4p}{(2\pi)^4}\e^{-\ui p\cdot(x-y)}\int\frac{\ud^4q}{(2\pi)^4}\tr_{l}\left[\gamma^\mu\frac{1}{\slashed p+\slashed q-\Sigma(-(p+q)^2)}\gamma^\nu\frac{1}{\slashed q-\Sigma(-q^2)}\right]\\
&=4\ui N_c\delta^{ij}\int\frac{\ud^4p}{(2\pi)^4}\e^{-\ui p\cdot(x-y)}\int\frac{\ud^4q}{(2\pi)^4}\\
&\times\left[-g^{\mu\nu}\Sigma(-(p+q)^2)\Sigma(-q^2)+g^{\mu\nu}p\cdot q+g^{\mu\nu}q^2-q^\mu p^\nu-p^\mu q^\nu-2q^\mu q^\nu\right]f(p\cdot q,q^2,p^2),
\end{aligned}
\end{equation}
where we shift integral momentum $p^\mu$ in the second equal sign, and 
\begin{equation}
f(p\cdot q,q^2,p^2)\equiv\frac{1}{(p+q)^2-\Sigma^2(-(p+q)^2)}\frac{1}{q^2-\Sigma^2(-q^2)}.
\end{equation}
We define some integral functions as
\begin{equation}
\begin{aligned}
\int\frac{\ud^4q}{(2\pi)^4}\Sigma(-(p+q)^2)\Sigma(-q^2) f(p\cdot q,q^2,p^2)&\equiv F_1(p^2),\\
\int\frac{\ud^4q}{(2\pi)^4}q^\mu f(p\cdot q,q^2,p^2)&\equiv p^\mu F_2(p^2),\\
\int\frac{\ud^4q}{(2\pi)^4}q^\mu q^\nu f(p\cdot q,q^2,p^2)&\equiv g^{\mu\nu}G_1(p^2)+p^\mu p^\nu G_2(p^2).
\end{aligned}
\end{equation}
Finally the vector current Green function becomes
\begin{equation}
\langle V_i^\mu(x)V_j^\nu(y)\rangle=4\ui N_c\delta^{ij}\int\frac{\ud^4p}{(2\pi)^4}\e^{-\ui p\cdot(x-y)}T^{\mu\nu},
\end{equation}
where 
\begin{equation}
T^{\mu\nu}\equiv\left[g^{\mu\nu}(-F_1+p^2F_2+2G_1+p^2G_2)-p^\mu p^\nu (2F_2+2G_2)\right].
\end{equation}
The conservation of vector current requires the following subtraction, as in Ref.~\cite{Bijnens:1995ww}:
\begin{equation}
T^{'\mu\nu}=T^{\mu\nu}-g^{\mu\nu}T^{\rho\sigma}\frac{p_\rho p_\sigma}{p^2}
\end{equation}
Then the conservation of vector current is satisfied: $p_\mu T^{'\mu\nu}=0$.
Compared to Eq.(\ref{eq:currentgf}), we obtain the spectral function $\rho_V(p^2)$:
\begin{equation}
\rho_V=-\frac{2\ui N_c}{\pi}p^2(2F_2+2G_2).
\end{equation}

Similarly, we can obtain the spectral function $\rho_A(p^2)$ as
\begin{equation}
\begin{aligned}
\rho_A=&-\frac{2\ui N_c}{\pi}(F_1+p^2F_2+2G_1+p^2G_2),\\
\frac{\rho_A}{p^2}+F_\pi^2\delta(p^2)=&-\frac{2\ui N_c}{\pi}(2F_2+2G_2)
\end{aligned}
\end{equation}
There is a relation between these two spectral functions:
\begin{equation}
\frac{\rho_V}{\mu^2}-\frac{\rho_A}{\mu^2}=F_\pi^2\delta(\mu^2).
\end{equation}
where $p^2=\mu^2$. After performing the integral, we obtain the Weinberg sum rules
\begin{equation}
\begin{aligned}
&\int_{0^-}^\infty[\rho_V(\mu^2)-\rho_A(\mu^2)]\mu^{-2}d\mu^{2}=\int_{0^-}^\infty F_\pi^2\delta(\mu^2)d\mu^{2}=F_\pi^2,\\
&\int_{0^-}^\infty[\rho_V(\mu^2)-\rho_A(\mu^2)]d\mu^{2}=\int_{0^-}^\infty F_\pi^2\mu^{2}\delta(\mu^2)d\mu^{2}=0.
\end{aligned}
\end{equation}

\section{Conclusion}

In this paper, we review our original derivation of effective chiral Lagrangian involving pseudo-scalar mesons and vector mesons (with hidden symmetry) from QCD. With proper approximations, we successfully obtain the analytical expressions for the $\mathcal O(p^2)$ and $\mathcal O(p^4)$ LECs in both the normal and anomalous parts of the chiral Lagrangian based on the quark self-energy. Using $|F_0^2|=(87\text{MeV})^2$ to fix the $\Lambda_{\text{QCD}}$ in the running coupling constant and choosing a cutoff $\Lambda=1000^{+100}_{-100}$MeV, we calculate the LECs numerically up to $\mathcal O(p^4)$. This marks the first time that LECs for vector mesons have been calculated directly from QCD, particularly for the critical parameters $a$ and $g$, which are typically treated as input values in the literature. The obtained $a$ is consistent with the phenomenological value, allowing us to reproduce many phenomenological results in our calculation. The obtained $g$ is slightly larger than the phenomenological value, but the non-Trln term in the action might compensate for this discrepancy \cite{Jiang:2015dba}. This aspect of the study is ongoing. The $\mathcal O(p^4)$ LECs we derive are comparable to those found in the literature, and we anticipate  more experiment values or model calculation for the $\mathcal O(p^4)$ LECs of HLS model in the future.

The methodology for deriving the chiral effective Lagrangian of vector mesons and calculating the LECs from QCD can be straightforwardly extended to axial-vector mesons, and such studies are also in progress.

In the studies focused on calculating the LECs in ChPT from underlying theory or corresponding approximate models, there are regularization dependence and sign issues for the $\mathcal O(p^2)$ LECs. 
The choice of regularization is generally guided by phenomenological reasons. Despite these challenges, we have demonstrated that our method is consistent with Weinberg sum rules. Further research is needed to address these issues comprehensively.

\section{Acknowledgments}
We would like to thank Shao-Zhou Jiang and Yong-Liang Ma for useful discussions.

\appendix{

\section{The $\mathcal O(p^4)$ Lagrangian in HLS model}\label{sec:hlsp4l}

The $\mathcal O(p^4)$ Lagrangian for general $N_f$ are given by \cite{Harada:2003jx}
\begin{equation}
\la_{(4)}=\la_{(4)y}+\la_{(4)\omega}+\la_{(4)z}\nonumber,
\end{equation}
where \begin{equation}
\begin{aligned}
\la_{(4)y}&=y_1\tr[\hat\alpha_{\bot\mu}\hat\alpha^\mu_\bot\hat\alpha_{\bot\nu}\hat\alpha^\nu_\bot]+y_2\tr[\hat\alpha_{\bot\mu}\hat\alpha_{\bot\nu}\hat\alpha^\mu_\bot\hat\alpha^\nu_\bot]+y_3\tr[\hat\alpha_{\parallel\mu}\hat\alpha^\mu_\parallel\hat\alpha_{\parallel\nu}\hat\alpha^\nu_\parallel]\\
&+y_4\tr[\hat\alpha_{\parallel\mu}\hat\alpha_{\parallel\nu}\hat\alpha^\mu_\parallel\hat\alpha^\nu_\parallel]+y_5\tr[\hat\alpha_{\bot\mu}\hat\alpha^\mu_\bot\hat\alpha_{\parallel\nu}\hat\alpha^\nu_\parallel]+y_6\tr[\hat\alpha_{\bot\mu}\hat\alpha_{\bot\nu}\hat\alpha^\mu_\parallel\hat\alpha^\nu_\parallel]\\
&+y_7\tr[\hat\alpha_{\bot\mu}\hat\alpha_{\bot\nu}\hat\alpha^\nu_\parallel\hat\alpha^\mu_\parallel]+y_8\{\tr[\hat\alpha_{\bot\mu}\hat\alpha^\mu_\parallel\hat\alpha_{\bot\nu}\hat\alpha^\nu_\parallel]+\tr[\hat\alpha_{\bot\mu}\hat\alpha_{\parallel\nu}\hat\alpha^\nu_\bot\hat\alpha^\mu_\parallel]\}\\
&+y_9\tr[\hat\alpha_{\bot\mu}\hat\alpha_{\parallel\nu}\hat\alpha^\mu_\bot\hat\alpha^\nu_\parallel]+y_{10}(\tr[\hat\alpha_{\bot\mu}\hat\alpha^\mu_\bot])^2+y_{11}\tr[\hat\alpha_{\bot\mu}\hat\alpha_{\bot\nu}]\tr[\hat\alpha^\mu_\bot\hat\alpha^\nu_\bot]\\
&+y_{12}(\tr[\hat\alpha_{\parallel\mu}\hat\alpha^\mu_\parallel])^2
+y_{13}\tr[\hat\alpha_{\parallel\mu}\hat\alpha_{\parallel\nu}]\tr[\hat\alpha^\mu_\parallel\hat\alpha^\nu_\parallel]+y_{14}\tr[\hat\alpha_{\bot\mu}\hat\alpha^\mu_\bot]\tr[\hat\alpha_{\parallel\nu}\hat\alpha^\nu_\parallel]\\
&+y_{15}\tr[\hat\alpha_{\bot\mu}\hat\alpha_{\bot\nu}]\tr[\hat\alpha^\mu_\parallel\hat\alpha^\nu_\parallel]+y_{16}(\tr[\hat\alpha_{\bot\mu}\hat\alpha^\mu_\parallel])^2+y_{17}\tr[\hat\alpha_{\bot\mu}\hat\alpha_{\parallel\nu}]\tr[\hat\alpha^\mu_\bot\hat\alpha^\nu_\parallel]\\
&+y_{18}\tr[\hat\alpha_{\bot\mu}\hat\alpha_{\parallel\nu}]\tr[\hat\alpha^\mu_\parallel\hat\alpha^\nu_\bot],
\end{aligned}
\end{equation}
\begin{equation}
\begin{aligned}
\la_{(4)\omega}&=\omega_1\tr[\hat\alpha_{\bot\mu}\hat\alpha^\mu_\bot(\hat\chi+\hat\chi^\dagger)]+\omega_2\tr[\hat\alpha_{\bot\mu}\hat\alpha^\mu_\bot]\tr[\hat\chi+\hat\chi^\dagger]\\
&+\omega_3\tr[\hat\alpha_{\parallel\mu}\hat\alpha^\mu_\parallel(\hat\chi+\hat\chi^\dagger)]+\omega_4\tr[\hat\alpha_{\parallel\mu}\hat\alpha^\mu_\parallel]\tr[\hat\chi+\hat\chi^\dagger]\\
&+\omega_5\tr[(\hat\alpha^\mu_\parallel\hat\alpha_{\bot\mu}-\hat\alpha_{\bot\mu}\hat\alpha^\mu_\parallel)(\hat\chi-\hat\chi^\dagger)]\\
&+\omega_6\tr[(\hat\chi+\hat\chi^\dagger)^2]+\omega_7(\tr[\hat\chi+\hat\chi^\dagger])^2\\
&+\omega_8\tr[(\hat\chi-\hat\chi^\dagger)^2]+\omega_9(\tr[\hat\chi-\hat\chi^\dagger])^2,
\end{aligned}
\end{equation}
\begin{equation}
\begin{aligned}
\la_{(4)z}&=z_1 \tr[\hat{\mathcal V}_{\mu\nu}\hat{\mathcal V}^{\mu\nu}]+z_2 \tr[\hat{\mathcal A}_{\mu\nu}\hat{\mathcal A}^{\mu\nu}]+z_3 \tr[\hat{\mathcal V}_{\mu\nu}V^{\mu\nu}]+\ui z_4\tr[V_{\mu\nu}\hat\alpha^\mu_\bot\hat\alpha^\nu_\bot]\\
&+\ui z_5\tr[V_{\mu\nu}\hat\alpha^\mu_\parallel\hat\alpha^\nu_\parallel]+\ui z_6\tr[\hat{\mathcal V}_{\mu\nu}\hat\alpha^\mu_\bot\hat\alpha^\nu_\bot]+\ui z_7\tr[\hat{\mathcal V}_{\mu\nu}\hat\alpha^\mu_\parallel\hat\alpha^\nu_\parallel]-\ui z_8 \tr[\hat{\mathcal A}_{\mu\nu}(\hat\alpha^\mu_\bot\hat\alpha^\nu_\parallel+\hat\alpha^\mu_\parallel\hat\alpha^\nu_\bot)]\\
&=z_1 \tr[\hat{\mathcal V}_{\mu\nu}\hat{\mathcal V}^{\mu\nu}]+z_2 \tr[\hat{\mathcal A}_{\mu\nu}\hat{\mathcal A}^{\mu\nu}]+z_3 \tr[\hat{\mathcal V}_{\mu\nu}V^{\mu\nu}]+\ui z_4\tr[V_{\mu\nu}a^\mu_\xi a^\nu_\xi]\\
&+\ui z_5\tr[V_{\mu\nu}{v'}^\mu_\xi {v'}^\nu_\xi]+\ui z_6\tr[\hat{\mathcal V}_{\mu\nu}a^\mu_\xi a^\nu_\xi]+\ui z_7\tr[\hat{\mathcal V}_{\mu\nu}{v'}^\mu_\xi {v'}^\nu_\xi]-\ui z_8 \tr[\hat{\mathcal A}_{\mu\nu}(a^\mu_\xi {v'}^\nu_\xi+{v'}^\mu_\xi a^\nu_\xi)],\label{eq:lp4z}
\end{aligned}
\end{equation}
where $\hat{\mathcal V}_{\mu\nu}\equiv\frac{1}{2}[\hat{\mathcal R}_{\mu\nu}+\hat{\mathcal L}_{\mu\nu}],~~\hat{\mathcal A}_{\mu\nu}\equiv\frac{1}{2}[\hat{\mathcal R}_{\mu\nu}-\hat{\mathcal L}_{\mu\nu}]$, and $\hat{\mathcal R}_{\mu\nu},\hat{\mathcal L}_{\mu\nu}$ are "converted" field strengths of the external gauge fields $\mathcal R_{\mu\nu},\mathcal L_{\mu\nu}$ respectively: 
\begin{equation}
\begin{aligned}
\hat{\mathcal R}_{\mu\nu}&=\xi_R\mathcal R_{\mu\nu}\xi_R^\dagger,~\hat{\mathcal L}_{\mu\nu}=\xi_L\mathcal L_{\mu\nu}\xi_L^\dagger.
\end{aligned}
\end{equation}

Integrating out the vector mesons by use of its equation of motion in the Lagrangian of the HLS, we obtain the $\mathcal O(p^4)$ Lagrangian for pseudoscalar mesons \cite{Gasser:1983yg,Gasser:1984gg} where the coefficients $L_i$ are expressed as the combinations of the coefficients in HLS model:
\begin{equation}
\begin{aligned}
L_1&=\frac{1}{32g^2}-\frac{1}{32}z_4+\frac{1}{32}y_2+\frac{1}{16}y_{10},\\
L_2&=\frac{1}{16g^2}-\frac{1}{16}z_4+\frac{1}{16}y_2+\frac{1}{16}y_{11},\\
L_3&=-\frac{3}{16g^2}+\frac{3}{16}z_4+\frac{1}{16}y_1-\frac{1}{8}y_{2},\\
L_4&=\frac{1}{4}\omega_2,~~~L_5=\frac{1}{4}\omega_1,~~~L_6=\omega_7,\\
L_7&=\omega_9,~~~L_8=\omega_6+\omega_8,\\
L_9&=\frac{1}{4}\left(\frac{1}{g^2}-z_3\right)-\frac{1}{8}(z_4+z_6),\\
L_{10}&=-\frac{1}{4g^2}+\frac{1}{2}(z_3-z_2+z_1),\\
H_1&=-\frac{1}{8g^2}+\frac{1}{4}(z_3+z_2+z_1),\\
H_2&=2(\omega_6-\omega_8).\label{eq:wyz2L}
\end{aligned}
\end{equation}

The $\mathcal O(p^4)$ anomalous Lagrangian in HLS model is \cite{Harada:2003jx}
\begin{equation}
\Gamma[\xi^\dagger_R\xi_L,V,l,r]=\Gamma_{\text{WZ}}[\xi^\dagger_R\xi_L,l,r]+\frac{N_c}{16\pi^2}\int\dif^4x\sum^4_{i=1}c_i\la_i,\label{eq:laap4hls}
\end{equation}
where $\Gamma_{\text{WZ}}$ is the well-known Wess-Zumino term, and $\la_i$ are gauge invariant terms which conserve parity and charge conjugation but violate the intrinsic parity:
\begin{equation}
\begin{aligned}
\la_1&=\ui\epsilon_{\mu\nu\sigma\rho}\tr_f[\hat\alpha_L^\mu\hat\alpha_L^\nu\hat\alpha_L^\sigma\hat\alpha_R^\rho-\hat\alpha_R^\mu\hat\alpha_R^\nu\hat\alpha_R^\sigma\hat\alpha_L^\rho],\\
\la_2&=\ui\epsilon_{\mu\nu\sigma\rho}\tr_f[\hat\alpha_L^\mu\hat\alpha_R^\nu\hat\alpha_L^\sigma\hat\alpha_R^\rho],\\
\la_3&=\epsilon_{\mu\nu\sigma\rho}\tr_f[F_V^{\mu\nu}(\hat\alpha_L^\sigma\hat\alpha_R^\rho-\hat\alpha_R^\sigma\hat\alpha_L^\rho)],\\
\la_4&=\frac{1}{2}\epsilon_{\mu\nu\sigma\rho}\tr_f[\hat F_L^{\mu\nu}(\hat\alpha_L^\sigma\hat\alpha_R^\rho-\hat\alpha_R^\sigma\hat\alpha_L^\rho)-\hat F_R^{\mu\nu}(\hat\alpha_R^\sigma\hat\alpha_L^\rho-\hat\alpha_L^\sigma\hat\alpha_R^\rho)],
\end{aligned}
\end{equation}
where the building blocks are given by
\begin{equation}
\begin{aligned}
&\hat\alpha_L^\mu=\hat\alpha^\mu_\parallel-\hat\alpha^\mu_\bot,~~\hat\alpha_R^\mu=\hat\alpha^\mu_\parallel+\hat\alpha^\mu_\bot,~~\epsilon_{\mu\nu\sigma\rho}F_V^{\mu\nu}=\frac{1}{2}\epsilon_{\mu\nu\sigma\rho}V^{\mu\nu},\\
&\epsilon_{\mu\nu\sigma\rho}\hat F_L^{\mu\nu}=\frac{1}{2}\epsilon_{\mu\nu\sigma\rho}(\hat{\mathcal V}^{\mu\nu}-\hat{\mathcal A}^{\mu\nu}),~~\epsilon_{\mu\nu\sigma\rho}\hat F_R^{\mu\nu}=\frac{1}{2}\epsilon_{\mu\nu\sigma\rho}(\hat{\mathcal V}^{\mu\nu}+\hat{\mathcal A}^{\mu\nu}).
\end{aligned}
\end{equation}

\section{The analytical expressions for $p^4$ order LECs}\label{sec:resp4}

\begin{equation}
\begin{aligned}
\C_2=&2\ui N_c\int\frac{\dif^4k}{(2\pi)^4}X^4[\Sigma^2k^2-2\Sigma^3{\Sigma'}k^2+\frac{2}{3}\Sigma{\Sigma'}k^4-\Sigma^4-\frac{1}{3}k^4-\frac{4}{3}\Sigma^2{\Sigma'}^2k^4]\\
&+\frac{\ui N_c}{2}\int\frac{\dif^4k}{(2\pi)^4}X^4[\frac{2}{3}k^4+\Sigma^4{\Sigma'}^2k^2+\frac{14}{3}\Sigma^2{\Sigma'}^2k^4+4\Sigma^3{\Sigma'}k^2+\frac{8}{3}\Sigma{\Sigma'}k^4+{\Sigma'}^2k^6+\Sigma^4]\\
&+\ui N_c\int\frac{\dif^4k}{(2\pi)^4}X^4[-\Sigma^2k^2-\Sigma^4{\Sigma'}^2k^2-5\Sigma^2{\Sigma'}^2k^4-4\Sigma^3{\Sigma'}k^2-\frac{8}{3}\Sigma{\Sigma'}k^4-{\Sigma'}^2k^6\\
&-{\Sigma}^4+\frac{1}{3}k^4+\frac{1}{3}{\Sigma'}^2k^6+(k^2-\Sigma^2)(2\Sigma'\Sigma k^2+\Sigma'\Sigma^3-\frac{7}{6}\Sigma''\Sigma k^4-\frac{1}{2}\Sigma''\Sigma^3 k^2)],
\end{aligned}
\end{equation}
\begin{equation}
\begin{aligned}
\C_3=&\frac{\ui N_c}{4}\int\frac{\dif^4k}{(2\pi)^4}X^4[\frac{2}{3}k^4+\Sigma^4{\Sigma'}^2k^2+\frac{14}{3}\Sigma^2{\Sigma'}^2k^4+4\Sigma^3{\Sigma'}k^2+\frac{8}{3}\Sigma{\Sigma'}k^4+{\Sigma'}^2k^6+\Sigma^4]\\
&+\frac{\ui N_c}{2}\int\frac{\dif^4k}{(2\pi)^4}X^4[-\Sigma^2k^2-\Sigma^4{\Sigma'}^2k^2-5\Sigma^2{\Sigma'}^2k^4-4\Sigma^3{\Sigma'}k^2-\frac{8}{3}\Sigma{\Sigma'}k^4-{\Sigma'}^2k^6\\
&-{\Sigma}^4+\frac{1}{3}k^4+\frac{1}{3}{\Sigma'}^2k^6+(k^2-\Sigma^2)(2\Sigma'\Sigma k^2+\Sigma'\Sigma^3-\frac{7}{6}\Sigma''\Sigma k^4-\frac{1}{2}\Sigma''\Sigma^3 k^2)],
\end{aligned}
\end{equation}
\begin{equation}
\begin{aligned}
\C_4&=2\ui N_c\int\frac{\dif^4k}{(2\pi)^4}\frac{\Sigma^4+3\Sigma^2k^2+\frac{1}{3}k^4}{[k^2-\Sigma^2(-k^2)]^4},\\
\C_5&=-\ui N_c\int\frac{\dif^4k}{(2\pi)^4}\frac{\Sigma^4+2\Sigma^2k^2+\frac{2}{3}k^4}{[k^2-\Sigma^2(-k^2)]^4},\\
\C_6&=2\ui N_c\int\frac{\dif^4k}{(2\pi)^4}\frac{k^2+\Sigma^2}{[k^2-\Sigma^2(-k^2)]^2},\\
\C_7&=2\ui N_c\int\frac{\dif^4k}{(2\pi)^4}\frac{1}{k^2-\Sigma^2(-k^2)},\\
\C_8&=-4\ui N_c\int\frac{\dif^4k}{(2\pi)^4}\frac{[2k^2+\Sigma^2]\Sigma}{[k^2-\Sigma^2(-k^2)]^3},
\end{aligned}
\end{equation}
\begin{align}
\C_9=&\frac{\ui N_c}{4}\int\frac{\dif^4k}{(2\pi)^4}\{[2\Sigma''\Sigma X]-4X^2\Sigma''k^2(k^2\Sigma'+\Sigma^2\Sigma'+\Sigma)\nonumber\\
&+2X^2({\Sigma'}^2-\Sigma'\Sigma''k^2+\Sigma''\Sigma''k^4/3)(k^2+\Sigma^2)\}+2X^3\{-2[\Sigma'\Sigma^3+\Sigma^3{\Sigma'}^3k^2\nonumber\\
&+3\Sigma^2{\Sigma'}^2k^2+3\Sigma{\Sigma'}^3k^4+{\Sigma'}^2k^4]\nonumber\\
&+[\Sigma^3\Sigma''k^2-\Sigma\Sigma''k^4]+\frac{4}{3}\Sigma''k^4[\Sigma'k^2(3\Sigma\Sigma'+1)+\Sigma(\Sigma^2{\Sigma'}^2+3\Sigma\Sigma'+1)]\}\nonumber\\
&-X^4[-2\Sigma^4-4{\Sigma'}^2\Sigma^4k^2-\frac{4}{3}{\Sigma'}^4\Sigma^4k^4-8\Sigma'\Sigma^3k^2-\frac{32}{3}{\Sigma'}^3\Sigma^3k^4-2\Sigma^2k^2\nonumber\\
&-\frac{40}{3}{\Sigma'}^2\Sigma^2k^4+\Sigma^2k^2-8{\Sigma'}^4\Sigma^2k^6-\frac{16}{3}\Sigma'\Sigma k^4+8\Sigma'\Sigma k^4-\frac{32}{3}{\Sigma'}^3\Sigma k^6\nonumber\\
&-2k^4+4{\Sigma'}^2k^6-\frac{4}{3}{\Sigma'}^4k^8-\frac{2}{3}k^4+2k^4-\frac{8}{3}{\Sigma'}^2k^6
]\}\nonumber\\
&-\frac{\ui N_c}{12}\int\frac{\dif^4k}{(2\pi)^4}\{X^2(\Sigma''\Sigma''k^4)(k^2+\Sigma^2)+4X^3\Sigma''k^4[\Sigma'k^2(3\Sigma\Sigma'+1)\nonumber\\
&+\Sigma(\Sigma^2{\Sigma'}^2+3\Sigma\Sigma'+1)]-3X^4[\Sigma^4-2{\Sigma'}^2\Sigma^4k^2-\frac{2}{3}{\Sigma'}^4\Sigma^4k^4-\frac{16}{3}{\Sigma'}^3\Sigma^3k^4\nonumber\\
&-\frac{20}{3}{\Sigma'}^2\Sigma^2k^4-\frac{1}{2}\Sigma^2k^2+4{\Sigma'}^2\Sigma^2k^4-4{\Sigma'}^4\Sigma^2k^6-\frac{8}{3}\Sigma'\Sigma k^4-\frac{16}{3}{\Sigma'}^3\Sigma k^6\nonumber\\
&+k^4-2{\Sigma'}^2k^6-\frac{2}{3}{\Sigma'}^4k^8-\frac{1}{3}k^4-\frac{4}{3}{\Sigma'}^2k^6]\},
\end{align}
\begin{equation}
\begin{aligned}
\C_{10}=&-4\ui N_c\int\frac{\dif^4k}{(2\pi)^4}X^4[-\frac{2}{3}k^4+2\Sigma^3{\Sigma'}k^2-\frac{2}{3}\Sigma{\Sigma'}k^4+{\Sigma}^4+\Sigma^2{\Sigma'}^2k^4+\frac{1}{3}{\Sigma'}^2k^6\\
&+\frac{1}{3}\Sigma''\Sigma k^6-\frac{1}{3}\Sigma''\Sigma^3 k^4]\\
&-\ui N_c\int\frac{\dif^4k}{(2\pi)^4}X^4[\frac{2}{3}k^4+\Sigma^4{\Sigma'}^2k^2+\frac{14}{3}\Sigma^2{\Sigma'}^2k^4+4\Sigma^3{\Sigma'}k^2+\frac{8}{3}\Sigma{\Sigma'}k^4+{\Sigma'}^2k^6+\Sigma^4]\\
&-2\ui N_c\int\frac{\dif^4k}{(2\pi)^4}X^4[-\Sigma^2k^2-\Sigma^4{\Sigma'}^2k^2-5\Sigma^2{\Sigma'}^2k^4-4\Sigma^3{\Sigma'}k^2-\frac{8}{3}\Sigma{\Sigma'}k^4-{\Sigma'}^2k^6\\
&-{\Sigma}^4+\frac{1}{3}k^4+\frac{1}{3}{\Sigma'}^2k^6+(k^2-\Sigma^2)(2\Sigma'\Sigma k^2+\Sigma'\Sigma^3-\frac{7}{6}\Sigma''\Sigma k^4-\frac{1}{2}\Sigma''\Sigma^3 k^2)],
\end{aligned}
\end{equation}
\begin{equation}
\C_{11}=-2\ui N_c\int\frac{\dif^4k}{(2\pi)^4}\frac{2\Sigma+k^2\Sigma'}{[k^2-\Sigma^2(-k^2)]^2},
\end{equation}
\begin{equation}
\begin{aligned}
\C_{12}=&2\ui N_c\int\frac{\dif^4k}{(2\pi)^4}X^4\{[\Sigma^2k^2-2\Sigma^3\Sigma'k^2+\frac{2}{3}\Sigma\Sigma'k^4-\Sigma^4-\frac{1}{3}k^4-\frac{4}{3}\Sigma^2{\Sigma'}^2k^4]\\
+&\frac{\ui N_c}{2}\int\frac{\dif^4k}{(2\pi)^4}X^4[\frac{2}{3}k^4-2\Sigma^2k^2-\Sigma^4{\Sigma'}^2k^2+\frac{2}{3}\Sigma^2{\Sigma'}^2k^4-{\Sigma'}^2k^6+\Sigma^4-\frac{4}{3}\Sigma{\Sigma'}k^4]\\
&-\frac{\ui N_c}{2}\int\frac{\dif^4k}{(2\pi)^4}X^4[\Sigma^2k^2-\Sigma^4{\Sigma'}^2k^2-2\Sigma^3\Sigma'k^2+\frac{2}{3}\Sigma\Sigma'k^4+\frac{2}{3}{\Sigma'}^2k^6-\Sigma^4\\
&-\frac{1}{3}k^4-\Sigma^2{\Sigma'}^2k^4+(k^2-\Sigma^2)(\Sigma^3\Sigma'+\frac{1}{2}\Sigma\Sigma''k^4-\frac{1}{2}\Sigma^3\Sigma''k^2-\frac{1}{3}\Sigma\Sigma''k^4)],
\end{aligned}
\end{equation}
\begin{equation}
\begin{aligned}
\C_{13}=&\frac{\ui N_c}{4}\int\frac{\dif^4k}{(2\pi)^4}X^4[\frac{2}{3}k^4-2\Sigma^2k^2-\Sigma^4{\Sigma'}^2k^2+\frac{2}{3}\Sigma^2{\Sigma'}^2k^4-{\Sigma'}^2k^6+\Sigma^4-\frac{4}{3}\Sigma{\Sigma'}k^4]\\
&-\frac{\ui N_c}{2}\int\frac{\dif^4k}{(2\pi)^4}X^4[\Sigma^2k^2-\Sigma^4{\Sigma'}^2k^2-2\Sigma^3\Sigma'k^2+\frac{2}{3}\Sigma\Sigma'k^4+\frac{2}{3}{\Sigma'}^2k^6-\Sigma^4\\
&-\frac{1}{3}k^4-\Sigma^2{\Sigma'}^2k^4+(k^2-\Sigma^2)(\Sigma^3\Sigma'+\frac{1}{2}\Sigma\Sigma''k^4-\frac{1}{2}\Sigma^3\Sigma''k^2-\frac{1}{3}\Sigma\Sigma''k^4)],
\end{aligned}
\end{equation}
\begin{align}
\C_{14}&=2\ui N_c\int\frac{\dif^4k}{(2\pi)^4}\frac{-\Sigma^2 k^2+\Sigma^4+\frac{1}{3}k^4}{[k^2-\Sigma^2(-k^2)]^4},\\
\C_{15}&=-\ui N_c\int\frac{\dif^4k}{(2\pi)^4}\frac{\frac{2}{3}k^4-2\Sigma^2 k^2+\Sigma^4}{[k^2-\Sigma^2(-k^2)]^4},\\
\C_{16}&=4\ui N_c\int\frac{\dif^4k}{(2\pi)^4}\frac{\Sigma^3}{[k^2-\Sigma^2(-k^2)]^3},
\end{align}
\begin{equation}
\begin{aligned}
\C_{17}=&-\ui N_c\int\frac{\dif^4k}{(2\pi)^4}X^4[\frac{2}{3}k^4-2\Sigma^2k^2-\Sigma^4{\Sigma'}^2k^2+\frac{2}{3}\Sigma^2{\Sigma'}^2k^4-{\Sigma'}^2k^6+\Sigma^4-\frac{4}{3}\Sigma{\Sigma'}k^4]\\
&-\frac{\ui N_c}{2}\int\frac{\dif^4k}{(2\pi)^4}X^4[\Sigma^2k^2-\Sigma^4{\Sigma'}^2k^2-2\Sigma^3\Sigma'k^2+\frac{2}{3}\Sigma\Sigma'k^4+\frac{2}{3}{\Sigma'}^2k^6-\Sigma^4\\
&-\frac{1}{3}k^4-\Sigma^2{\Sigma'}^2k^4+(k^2-\Sigma^2)(\Sigma^3\Sigma'+\frac{1}{2}\Sigma\Sigma''k^4-\frac{1}{2}\Sigma^3\Sigma''k^2-\frac{1}{3}\Sigma\Sigma''k^4)]\\
&+4\ui N_c\int\frac{\dif^4k}{(2\pi)^4}X^4[\frac{2}{3}k^4-2\Sigma^2k^2+\Sigma^4-\Sigma^2{\Sigma'}^2k^4-\frac{4}{3}\Sigma\Sigma'k^4-\frac{1}{3}{\Sigma'}^2k^6\\
&-(k^2-\Sigma^2)(\frac{1}{3}\Sigma\Sigma''k^4)],
\end{aligned}
\end{equation}

\begin{equation}
\begin{aligned}
\C_{19}&=4\ui N_c\int\frac{\dif^4k}{(2\pi)^4}\frac{-\Sigma^2 k^2-\Sigma^4+\frac{1}{3}k^4}{[k^2-\Sigma^2(-k^2)]^4},\\
\C_{20}&=4\ui N_c\int\frac{\dif^4k}{(2\pi)^4}\frac{-\frac{2}{3}k^4+\Sigma^4}{[k^2-\Sigma^2(-k^2)]^4},\\
\C_{21}&=4\ui N_c\int\frac{\dif^4k}{(2\pi)^4}\frac{\Sigma^2 k^2-\Sigma^4+\frac{1}{3}k^4}{[k^2-\Sigma^2(-k^2)]^4},\\
\C_{22}&=-2\ui N_c\int\frac{\dif^4k}{(2\pi)^4}\frac{-\frac{1}{3}k^4+\Sigma^2k^2-\Sigma^4}{[k^2-\Sigma^2(-k^2)]^4},\\
\C_{23}&=-2\ui N_c\int\frac{\dif^4k}{(2\pi)^4}\frac{\frac{2}{3}k^4+\Sigma^4}{[k^2-\Sigma^2(-k^2)]^4},\\
\C_{25}&=4N_c\int\frac{\dif^4k}{(2\pi)^4}\frac{\Sigma}{[k^2-\Sigma^2(-k^2)]^2},
\end{aligned}
\end{equation}
\begin{equation}
\begin{aligned}
\C_{27}=&-4N_c\int\frac{\dif^4k}{(2\pi)^4}X^4[-\frac{2}{3}k^4+\Sigma^3\Sigma'k^2-\frac{1}{3}\Sigma\Sigma'k^4+\Sigma^4]\\
&-2N_c\int\frac{\dif^4k}{(2\pi)^4}X^4\{[-\Sigma^2k^2-2\Sigma^3\Sigma'k^2-\frac{4}{3}\Sigma\Sigma'k^4-\Sigma^4+\frac{1}{3}k^4]\\
&+\frac{1}{2}\times[\frac{2}{3}k^4+2\Sigma^3\Sigma'k^2+\frac{4}{3}\Sigma\Sigma'k^4+\Sigma^4]\},
\end{aligned}
\end{equation}
\begin{align}
\C_{28}=&\ui N_c\int\frac{\dif^4k}{(2\pi)^4}X^4\{-1\times[-\frac{1}{3}k^4-\Sigma^4{\Sigma'}^2k^2-\frac{4}{3}\Sigma^3{\Sigma'}^3k^4-3\Sigma^3\Sigma'k^2-\frac{10}{3}\Sigma^2{\Sigma'}^2k^4\nonumber\\
&-\frac{4}{3}\Sigma{\Sigma'}^3k^6+\frac{1}{3}{\Sigma'}^2k^6+\Sigma\Sigma'k^4+\Sigma^2k^2-\Sigma^4+(k^2-\Sigma^2)(\Sigma^3\Sigma'+\frac{3}{2}\Sigma^2{\Sigma'}^2k^2\nonumber\\
&+\frac{1}{2}{\Sigma'}^2k^4-\frac{1}{6}\Sigma\Sigma''k^4-\frac{1}{2}\Sigma^3\Sigma''k^2-\Sigma^2\Sigma'\Sigma''k^4-\frac{1}{3}\Sigma'\Sigma''k^6)+\frac{1}{2}\Sigma\Sigma''(k^2-\Sigma^2)^2]\nonumber\\
&+\frac{1}{2}\times[\frac{2}{3}k^4-\Sigma^4{\Sigma'}^2k^2-\frac{4}{3}\Sigma^3{\Sigma'}^3k^4-\frac{4}{3}\Sigma^2{\Sigma'}^2k^4-\frac{5}{3}{\Sigma'}^2k^6-2\Sigma\Sigma'k^4\nonumber\\
&-\frac{4}{3}\Sigma{\Sigma'}^3k^6-2\Sigma^2k^2+\Sigma^4-(k^2-\Sigma^2)(\Sigma^2\Sigma'\Sigma''k^4+\frac{2}{3}\Sigma
\Sigma''k^4+\frac{1}{3}\Sigma'\Sigma''k^6)],
\end{align}
\begin{equation}
\begin{aligned}
\C_{29}=&2N_c\int\frac{\dif^4k}{(2\pi)^4}X^4\{[-\Sigma^2k^2-2\Sigma^3\Sigma'k^2-\frac{4}{3}\Sigma\Sigma'k^4-\Sigma^4+\frac{1}{3}k^4]\\
&+\frac{1}{2}\times[\frac{2}{3}k^4+2\Sigma^3\Sigma'k^2+\frac{4}{3}\Sigma\Sigma'k^4+\Sigma^4]\},
\end{aligned}
\end{equation}
\begin{equation}
\begin{aligned}
\C_{31}=&-4N_c\int\frac{\dif^4k}{(2\pi)^4}X^4[\Sigma^2k^2-\Sigma^3\Sigma'k^2+\frac{1}{3}\Sigma\Sigma'k^4-\Sigma^4-\frac{1}{3}k^4]\\
&-2N_c\int\frac{\dif^4k}{(2\pi)^4}X^4\{[-\Sigma^2k^2-2\Sigma^3\Sigma'k^2-\frac{4}{3}\Sigma\Sigma'k^4-\Sigma^4+\frac{1}{3}k^4]\\
&+\frac{1}{2}\times[\frac{2}{3}k^4+2\Sigma^3\Sigma'k^2+\frac{4}{3}\Sigma\Sigma'k^4+\Sigma^4]\}.
\end{aligned}
\end{equation}

}


\bibliography{vectorLECs.bib}

\begin{thebibliography}{54}
\expandafter\ifx\csname natexlab\endcsname\relax\def\natexlab#1{#1}\fi
\expandafter\ifx\csname bibnamefont\endcsname\relax
  \def\bibnamefont#1{#1}\fi
\expandafter\ifx\csname bibfnamefont\endcsname\relax
  \def\bibfnamefont#1{#1}\fi
\expandafter\ifx\csname citenamefont\endcsname\relax
  \def\citenamefont#1{#1}\fi
\expandafter\ifx\csname url\endcsname\relax
  \def\url#1{\texttt{#1}}\fi
\expandafter\ifx\csname urlprefix\endcsname\relax\def\urlprefix{URL }\fi
\providecommand{\bibinfo}[2]{#2}
\providecommand{\eprint}[2][]{\url{#2}}

\bibitem[{\citenamefont{Weinberg}(1979)}]{Weinberg:1978kz}
\bibinfo{author}{\bibfnamefont{S.}~\bibnamefont{Weinberg}},
  \bibinfo{journal}{Physica A} \textbf{\bibinfo{volume}{96}},
  \bibinfo{pages}{327} (\bibinfo{year}{1979}).

\bibitem[{\citenamefont{Gasser and Leutwyler}(1984)}]{Gasser:1983yg}
\bibinfo{author}{\bibfnamefont{J.}~\bibnamefont{Gasser}} \bibnamefont{and}
  \bibinfo{author}{\bibfnamefont{H.}~\bibnamefont{Leutwyler}},
  \bibinfo{journal}{Annals Phys.} \textbf{\bibinfo{volume}{158}},
  \bibinfo{pages}{142} (\bibinfo{year}{1984}).

\bibitem[{\citenamefont{Gasser and Leutwyler}(1985)}]{Gasser:1984gg}
\bibinfo{author}{\bibfnamefont{J.}~\bibnamefont{Gasser}} \bibnamefont{and}
  \bibinfo{author}{\bibfnamefont{H.}~\bibnamefont{Leutwyler}},
  \bibinfo{journal}{Nucl. Phys. B} \textbf{\bibinfo{volume}{250}},
  \bibinfo{pages}{465} (\bibinfo{year}{1985}).

\bibitem[{\citenamefont{Bijnens et~al.}(2019)\citenamefont{Bijnens,
  Hermansson-Truedsson, and Wang}}]{Bijnens:2018lez}
\bibinfo{author}{\bibfnamefont{J.}~\bibnamefont{Bijnens}},
  \bibinfo{author}{\bibfnamefont{N.}~\bibnamefont{Hermansson-Truedsson}},
  \bibnamefont{and} \bibinfo{author}{\bibfnamefont{S.}~\bibnamefont{Wang}},
  \bibinfo{journal}{JHEP} \textbf{\bibinfo{volume}{01}}, \bibinfo{pages}{102}
  (\bibinfo{year}{2019}), \eprint{1810.06834}.

\bibitem[{\citenamefont{Ecker et~al.}(1989{\natexlab{a}})\citenamefont{Ecker,
  Gasser, Leutwyler, Pich, and de~Rafael}}]{Ecker:1989yg}
\bibinfo{author}{\bibfnamefont{G.}~\bibnamefont{Ecker}},
  \bibinfo{author}{\bibfnamefont{J.}~\bibnamefont{Gasser}},
  \bibinfo{author}{\bibfnamefont{H.}~\bibnamefont{Leutwyler}},
  \bibinfo{author}{\bibfnamefont{A.}~\bibnamefont{Pich}}, \bibnamefont{and}
  \bibinfo{author}{\bibfnamefont{E.}~\bibnamefont{de~Rafael}},
  \bibinfo{journal}{Phys. Lett. B} \textbf{\bibinfo{volume}{223}},
  \bibinfo{pages}{425} (\bibinfo{year}{1989}{\natexlab{a}}).

\bibitem[{\citenamefont{Schwinger}(1967)}]{Schwinger:1967tc}
\bibinfo{author}{\bibfnamefont{J.~S.} \bibnamefont{Schwinger}},
  \bibinfo{journal}{Phys. Lett. B} \textbf{\bibinfo{volume}{24}},
  \bibinfo{pages}{473} (\bibinfo{year}{1967}).

\bibitem[{\citenamefont{Kaymakcalan and Schechter}(1985)}]{Kaymakcalan:1984bz}
\bibinfo{author}{\bibfnamefont{O.}~\bibnamefont{Kaymakcalan}} \bibnamefont{and}
  \bibinfo{author}{\bibfnamefont{J.}~\bibnamefont{Schechter}},
  \bibinfo{journal}{Phys. Rev. D} \textbf{\bibinfo{volume}{31}},
  \bibinfo{pages}{1109} (\bibinfo{year}{1985}).

\bibitem[{\citenamefont{Meissner}(1988)}]{Meissner:1987ge}
\bibinfo{author}{\bibfnamefont{U.~G.} \bibnamefont{Meissner}},
  \bibinfo{journal}{Phys. Rept.} \textbf{\bibinfo{volume}{161}},
  \bibinfo{pages}{213} (\bibinfo{year}{1988}).

\bibitem[{\citenamefont{Bando et~al.}(1988)\citenamefont{Bando, Kugo, and
  Yamawaki}}]{Bando:1987br}
\bibinfo{author}{\bibfnamefont{M.}~\bibnamefont{Bando}},
  \bibinfo{author}{\bibfnamefont{T.}~\bibnamefont{Kugo}}, \bibnamefont{and}
  \bibinfo{author}{\bibfnamefont{K.}~\bibnamefont{Yamawaki}},
  \bibinfo{journal}{Phys. Rept.} \textbf{\bibinfo{volume}{164}},
  \bibinfo{pages}{217} (\bibinfo{year}{1988}).

\bibitem[{\citenamefont{Harada and Yamawaki}(2003)}]{Harada:2003jx}
\bibinfo{author}{\bibfnamefont{M.}~\bibnamefont{Harada}} \bibnamefont{and}
  \bibinfo{author}{\bibfnamefont{K.}~\bibnamefont{Yamawaki}},
  \bibinfo{journal}{Phys. Rept.} \textbf{\bibinfo{volume}{381}},
  \bibinfo{pages}{1} (\bibinfo{year}{2003}), \eprint{hep-ph/0302103}.

\bibitem[{\citenamefont{Ecker et~al.}(1989{\natexlab{b}})\citenamefont{Ecker,
  Gasser, Pich, and de~Rafael}}]{Ecker:1988te}
\bibinfo{author}{\bibfnamefont{G.}~\bibnamefont{Ecker}},
  \bibinfo{author}{\bibfnamefont{J.}~\bibnamefont{Gasser}},
  \bibinfo{author}{\bibfnamefont{A.}~\bibnamefont{Pich}}, \bibnamefont{and}
  \bibinfo{author}{\bibfnamefont{E.}~\bibnamefont{de~Rafael}},
  \bibinfo{journal}{Nucl. Phys. B} \textbf{\bibinfo{volume}{321}},
  \bibinfo{pages}{311} (\bibinfo{year}{1989}{\natexlab{b}}).

\bibitem[{\citenamefont{Tanabashi}(1996)}]{Tanabashi:1995nz}
\bibinfo{author}{\bibfnamefont{M.}~\bibnamefont{Tanabashi}},
  \bibinfo{journal}{Phys. Lett. B} \textbf{\bibinfo{volume}{384}},
  \bibinfo{pages}{218} (\bibinfo{year}{1996}), \eprint{hep-ph/9511367}.

\bibitem[{\citenamefont{Graf et~al.}(2021)\citenamefont{Graf, Henning, Lu,
  Melia, and Murayama}}]{Graf:2020yxt}
\bibinfo{author}{\bibfnamefont{L.}~\bibnamefont{Graf}},
  \bibinfo{author}{\bibfnamefont{B.}~\bibnamefont{Henning}},
  \bibinfo{author}{\bibfnamefont{X.}~\bibnamefont{Lu}},
  \bibinfo{author}{\bibfnamefont{T.}~\bibnamefont{Melia}}, \bibnamefont{and}
  \bibinfo{author}{\bibfnamefont{H.}~\bibnamefont{Murayama}},
  \bibinfo{journal}{JHEP} \textbf{\bibinfo{volume}{01}}, \bibinfo{pages}{142}
  (\bibinfo{year}{2021}), \eprint{2009.01239}.

\bibitem[{\citenamefont{Guo et~al.}(2020)\citenamefont{Guo, Yang, and
  Jiang}}]{Guo:2020cmv}
\bibinfo{author}{\bibfnamefont{W.}~\bibnamefont{Guo}},
  \bibinfo{author}{\bibfnamefont{Q.-H.} \bibnamefont{Yang}}, \bibnamefont{and}
  \bibinfo{author}{\bibfnamefont{S.-Z.} \bibnamefont{Jiang}}
  (\bibinfo{year}{2020}), \eprint{2006.15258}.

\bibitem[{\citenamefont{Amoros et~al.}(2000)\citenamefont{Amoros, Bijnens, and
  Talavera}}]{Amoros:1999dp}
\bibinfo{author}{\bibfnamefont{G.}~\bibnamefont{Amoros}},
  \bibinfo{author}{\bibfnamefont{J.}~\bibnamefont{Bijnens}}, \bibnamefont{and}
  \bibinfo{author}{\bibfnamefont{P.}~\bibnamefont{Talavera}},
  \bibinfo{journal}{Nucl. Phys. B} \textbf{\bibinfo{volume}{568}},
  \bibinfo{pages}{319} (\bibinfo{year}{2000}), \eprint{hep-ph/9907264}.

\bibitem[{\citenamefont{Knecht and Nyffeler}(2001)}]{Knecht:2001xc}
\bibinfo{author}{\bibfnamefont{M.}~\bibnamefont{Knecht}} \bibnamefont{and}
  \bibinfo{author}{\bibfnamefont{A.}~\bibnamefont{Nyffeler}},
  \bibinfo{journal}{Eur. Phys. J. C} \textbf{\bibinfo{volume}{21}},
  \bibinfo{pages}{659} (\bibinfo{year}{2001}), \eprint{hep-ph/0106034}.

\bibitem[{\citenamefont{Golterman et~al.}(2014)\citenamefont{Golterman,
  Maltman, and Peris}}]{Golterman:2014nua}
\bibinfo{author}{\bibfnamefont{M.}~\bibnamefont{Golterman}},
  \bibinfo{author}{\bibfnamefont{K.}~\bibnamefont{Maltman}}, \bibnamefont{and}
  \bibinfo{author}{\bibfnamefont{S.}~\bibnamefont{Peris}},
  \bibinfo{journal}{Phys. Rev. D} \textbf{\bibinfo{volume}{89}},
  \bibinfo{pages}{054036} (\bibinfo{year}{2014}), \eprint{1402.1043}.

\bibitem[{\citenamefont{Necco}(2008)}]{Necco:2008ish}
\bibinfo{author}{\bibfnamefont{S.}~\bibnamefont{Necco}}, \bibinfo{journal}{PoS}
  \textbf{\bibinfo{volume}{CONFINEMENT8}}, \bibinfo{pages}{024}
  (\bibinfo{year}{2008}), \eprint{0901.4257}.

\bibitem[{\citenamefont{Aoki et~al.}(2017)}]{Aoki:2016frl}
\bibinfo{author}{\bibfnamefont{S.}~\bibnamefont{Aoki}} \bibnamefont{et~al.},
  \bibinfo{journal}{Eur. Phys. J. C} \textbf{\bibinfo{volume}{77}},
  \bibinfo{pages}{112} (\bibinfo{year}{2017}), \eprint{1607.00299}.

\bibitem[{\citenamefont{Harada et~al.}(2006)\citenamefont{Harada, Matsuzaki,
  and Yamawaki}}]{Harada:2006di}
\bibinfo{author}{\bibfnamefont{M.}~\bibnamefont{Harada}},
  \bibinfo{author}{\bibfnamefont{S.}~\bibnamefont{Matsuzaki}},
  \bibnamefont{and} \bibinfo{author}{\bibfnamefont{K.}~\bibnamefont{Yamawaki}},
  \bibinfo{journal}{Phys. Rev. D} \textbf{\bibinfo{volume}{74}},
  \bibinfo{pages}{076004} (\bibinfo{year}{2006}), \eprint{hep-ph/0603248}.

\bibitem[{\citenamefont{Harada et~al.}(2010)\citenamefont{Harada, Matsuzaki,
  and Yamawaki}}]{Harada:2010cn}
\bibinfo{author}{\bibfnamefont{M.}~\bibnamefont{Harada}},
  \bibinfo{author}{\bibfnamefont{S.}~\bibnamefont{Matsuzaki}},
  \bibnamefont{and} \bibinfo{author}{\bibfnamefont{K.}~\bibnamefont{Yamawaki}},
  \bibinfo{journal}{Phys. Rev. D} \textbf{\bibinfo{volume}{82}},
  \bibinfo{pages}{076010} (\bibinfo{year}{2010}), \eprint{1007.4715}.

\bibitem[{\citenamefont{Ma et~al.}(2013)\citenamefont{Ma, Yang, Oh, and
  Harada}}]{Ma:2012zm}
\bibinfo{author}{\bibfnamefont{Y.-L.} \bibnamefont{Ma}},
  \bibinfo{author}{\bibfnamefont{G.-S.} \bibnamefont{Yang}},
  \bibinfo{author}{\bibfnamefont{Y.}~\bibnamefont{Oh}}, \bibnamefont{and}
  \bibinfo{author}{\bibfnamefont{M.}~\bibnamefont{Harada}},
  \bibinfo{journal}{Phys. Rev. D} \textbf{\bibinfo{volume}{87}},
  \bibinfo{pages}{034023} (\bibinfo{year}{2013}), \eprint{1209.3554}.

\bibitem[{\citenamefont{Wang et~al.}(2000)\citenamefont{Wang, Kuang, Xiao, and
  Wang}}]{Wang:1999cp}
\bibinfo{author}{\bibfnamefont{Q.}~\bibnamefont{Wang}},
  \bibinfo{author}{\bibfnamefont{Y.-P.} \bibnamefont{Kuang}},
  \bibinfo{author}{\bibfnamefont{M.}~\bibnamefont{Xiao}}, \bibnamefont{and}
  \bibinfo{author}{\bibfnamefont{X.-L.} \bibnamefont{Wang}},
  \bibinfo{journal}{Phys. Rev. D} \textbf{\bibinfo{volume}{61}},
  \bibinfo{pages}{054011} (\bibinfo{year}{2000}), \eprint{hep-ph/9903201}.

\bibitem[{\citenamefont{Yang et~al.}(2002)\citenamefont{Yang, Wang, Kuang, and
  Lu}}]{Yang:2002hea}
\bibinfo{author}{\bibfnamefont{H.}~\bibnamefont{Yang}},
  \bibinfo{author}{\bibfnamefont{Q.}~\bibnamefont{Wang}},
  \bibinfo{author}{\bibfnamefont{Y.-P.} \bibnamefont{Kuang}}, \bibnamefont{and}
  \bibinfo{author}{\bibfnamefont{Q.}~\bibnamefont{Lu}}, \bibinfo{journal}{Phys.
  Rev. D} \textbf{\bibinfo{volume}{66}}, \bibinfo{pages}{014019}
  (\bibinfo{year}{2002}), \eprint{hep-ph/0203040}.

\bibitem[{\citenamefont{Jiang et~al.}(2010)\citenamefont{Jiang, Zhang, Li, and
  Wang}}]{Jiang:2009uf}
\bibinfo{author}{\bibfnamefont{S.-Z.} \bibnamefont{Jiang}},
  \bibinfo{author}{\bibfnamefont{Y.}~\bibnamefont{Zhang}},
  \bibinfo{author}{\bibfnamefont{C.}~\bibnamefont{Li}}, \bibnamefont{and}
  \bibinfo{author}{\bibfnamefont{Q.}~\bibnamefont{Wang}},
  \bibinfo{journal}{Phys. Rev. D} \textbf{\bibinfo{volume}{81}},
  \bibinfo{pages}{014001} (\bibinfo{year}{2010}), \eprint{0907.5229}.

\bibitem[{\citenamefont{Jiang and Wang}(2010)}]{Jiang:2010wa}
\bibinfo{author}{\bibfnamefont{S.-Z.} \bibnamefont{Jiang}} \bibnamefont{and}
  \bibinfo{author}{\bibfnamefont{Q.}~\bibnamefont{Wang}},
  \bibinfo{journal}{Phys. Rev. D} \textbf{\bibinfo{volume}{81}},
  \bibinfo{pages}{094037} (\bibinfo{year}{2010}), \eprint{1001.0315}.

\bibitem[{\citenamefont{Bijnens and Jemos}(2012)}]{Bijnens:2011tb}
\bibinfo{author}{\bibfnamefont{J.}~\bibnamefont{Bijnens}} \bibnamefont{and}
  \bibinfo{author}{\bibfnamefont{I.}~\bibnamefont{Jemos}},
  \bibinfo{journal}{Nucl. Phys. B} \textbf{\bibinfo{volume}{854}},
  \bibinfo{pages}{631} (\bibinfo{year}{2012}), \eprint{1103.5945}.

\bibitem[{\citenamefont{Wang and Wang}(2000)}]{Wang:2000mu}
\bibinfo{author}{\bibfnamefont{X.-L.} \bibnamefont{Wang}} \bibnamefont{and}
  \bibinfo{author}{\bibfnamefont{Q.}~\bibnamefont{Wang}},
  \bibinfo{journal}{Commun. Theor. Phys.} \textbf{\bibinfo{volume}{34}},
  \bibinfo{pages}{519} (\bibinfo{year}{2000}).

\bibitem[{\citenamefont{Balog}(1984)}]{Balog:1984upv}
\bibinfo{author}{\bibfnamefont{J.}~\bibnamefont{Balog}},
  \bibinfo{journal}{Phys. Lett. B} \textbf{\bibinfo{volume}{149}},
  \bibinfo{pages}{197} (\bibinfo{year}{1984}).

\bibitem[{\citenamefont{Andrianov}(1985)}]{Andrianov:1985ay}
\bibinfo{author}{\bibfnamefont{A.~A.} \bibnamefont{Andrianov}},
  \bibinfo{journal}{Phys. Lett. B} \textbf{\bibinfo{volume}{157}},
  \bibinfo{pages}{425} (\bibinfo{year}{1985}).

\bibitem[{\citenamefont{Klevansky}(1992)}]{Klevansky:1992qe}
\bibinfo{author}{\bibfnamefont{S.~P.} \bibnamefont{Klevansky}},
  \bibinfo{journal}{Rev. Mod. Phys.} \textbf{\bibinfo{volume}{64}},
  \bibinfo{pages}{649} (\bibinfo{year}{1992}).

\bibitem[{\citenamefont{Hatsuda and Kunihiro}(1994)}]{Hatsuda:1994pi}
\bibinfo{author}{\bibfnamefont{T.}~\bibnamefont{Hatsuda}} \bibnamefont{and}
  \bibinfo{author}{\bibfnamefont{T.}~\bibnamefont{Kunihiro}},
  \bibinfo{journal}{Phys. Rept.} \textbf{\bibinfo{volume}{247}},
  \bibinfo{pages}{221} (\bibinfo{year}{1994}), \eprint{hep-ph/9401310}.

\bibitem[{\citenamefont{Bijnens}(1996)}]{Bijnens:1995ww}
\bibinfo{author}{\bibfnamefont{J.}~\bibnamefont{Bijnens}},
  \bibinfo{journal}{Phys. Rept.} \textbf{\bibinfo{volume}{265}},
  \bibinfo{pages}{369} (\bibinfo{year}{1996}), \eprint{hep-ph/9502335}.

\bibitem[{\citenamefont{Holdom}(1992)}]{Holdom:1992fn}
\bibinfo{author}{\bibfnamefont{B.}~\bibnamefont{Holdom}},
  \bibinfo{journal}{Phys. Rev. D} \textbf{\bibinfo{volume}{45}},
  \bibinfo{pages}{2534} (\bibinfo{year}{1992}).

\bibitem[{\citenamefont{Jiang et~al.}(2015)\citenamefont{Jiang, Wei, Chen, and
  Wang}}]{Jiang:2015dba}
\bibinfo{author}{\bibfnamefont{S.-Z.} \bibnamefont{Jiang}},
  \bibinfo{author}{\bibfnamefont{Z.-L.} \bibnamefont{Wei}},
  \bibinfo{author}{\bibfnamefont{Q.-S.} \bibnamefont{Chen}}, \bibnamefont{and}
  \bibinfo{author}{\bibfnamefont{Q.}~\bibnamefont{Wang}},
  \bibinfo{journal}{Phys. Rev. D} \textbf{\bibinfo{volume}{92}},
  \bibinfo{pages}{025014} (\bibinfo{year}{2015}), \eprint{1502.05087}.

\bibitem[{\citenamefont{Lu et~al.}(2002)\citenamefont{Lu, Yang, and
  Wang}}]{Lu:2002fx}
\bibinfo{author}{\bibfnamefont{Q.}~\bibnamefont{Lu}},
  \bibinfo{author}{\bibfnamefont{H.}~\bibnamefont{Yang}}, \bibnamefont{and}
  \bibinfo{author}{\bibfnamefont{Q.}~\bibnamefont{Wang}},
  \bibinfo{journal}{Commun. Theor. Phys.} \textbf{\bibinfo{volume}{38}},
  \bibinfo{pages}{185} (\bibinfo{year}{2002}), \eprint{hep-ph/0202078}.

\bibitem[{\citenamefont{Espriu et~al.}(1990)\citenamefont{Espriu, de~Rafael,
  and Taron}}]{Espriu:1989ff}
\bibinfo{author}{\bibfnamefont{D.}~\bibnamefont{Espriu}},
  \bibinfo{author}{\bibfnamefont{E.}~\bibnamefont{de~Rafael}},
  \bibnamefont{and} \bibinfo{author}{\bibfnamefont{J.}~\bibnamefont{Taron}},
  \bibinfo{journal}{Nucl. Phys. B} \textbf{\bibinfo{volume}{345}},
  \bibinfo{pages}{22} (\bibinfo{year}{1990}), \bibinfo{note}{[Erratum:
  Nucl.Phys.B 355, 278--279 (1991)]}.

\bibitem[{\citenamefont{Chen et~al.}(2020)\citenamefont{Chen, Fu, Ma, and
  Wang}}]{Chen:2020jiq}
\bibinfo{author}{\bibfnamefont{Q.-S.} \bibnamefont{Chen}},
  \bibinfo{author}{\bibfnamefont{H.-F.} \bibnamefont{Fu}},
  \bibinfo{author}{\bibfnamefont{Y.-L.} \bibnamefont{Ma}}, \bibnamefont{and}
  \bibinfo{author}{\bibfnamefont{Q.}~\bibnamefont{Wang}},
  \bibinfo{journal}{Phys. Rev. D} \textbf{\bibinfo{volume}{102}},
  \bibinfo{pages}{034034} (\bibinfo{year}{2020}), \eprint{2001.06418}.

\bibitem[{\citenamefont{Chen et~al.}(2021)\citenamefont{Chen, Fu, Ma, and
  Wang}}]{Chen:2021qkx}
\bibinfo{author}{\bibfnamefont{Q.-S.} \bibnamefont{Chen}},
  \bibinfo{author}{\bibfnamefont{H.-F.} \bibnamefont{Fu}},
  \bibinfo{author}{\bibfnamefont{Y.-L.} \bibnamefont{Ma}}, \bibnamefont{and}
  \bibinfo{author}{\bibfnamefont{Q.}~\bibnamefont{Wang}},
  \bibinfo{journal}{Commun. Theor. Phys.} \textbf{\bibinfo{volume}{73}},
  \bibinfo{pages}{065202} (\bibinfo{year}{2021}), \eprint{2101.07105}.

\bibitem[{\citenamefont{Weinberg}(1967)}]{Weinberg:1967kj}
\bibinfo{author}{\bibfnamefont{S.}~\bibnamefont{Weinberg}},
  \bibinfo{journal}{Phys. Rev. Lett.} \textbf{\bibinfo{volume}{18}},
  \bibinfo{pages}{507} (\bibinfo{year}{1967}).

\bibitem[{\citenamefont{Kawarabayashi and Suzuki}(1966)}]{Kawarabayashi:1966kd}
\bibinfo{author}{\bibfnamefont{K.}~\bibnamefont{Kawarabayashi}}
  \bibnamefont{and} \bibinfo{author}{\bibfnamefont{M.}~\bibnamefont{Suzuki}},
  \bibinfo{journal}{Phys. Rev. Lett.} \textbf{\bibinfo{volume}{16}},
  \bibinfo{pages}{255} (\bibinfo{year}{1966}).

\bibitem[{\citenamefont{Riazuddin and Fayyazuddin}(1966)}]{Riazuddin:1966sw}
\bibinfo{author}{\bibnamefont{Riazuddin}} \bibnamefont{and}
  \bibinfo{author}{\bibnamefont{Fayyazuddin}}, \bibinfo{journal}{Phys. Rev.}
  \textbf{\bibinfo{volume}{147}}, \bibinfo{pages}{1071} (\bibinfo{year}{1966}).

\bibitem[{\citenamefont{Pagels and Stokar}(1979)}]{Pagels:1979hd}
\bibinfo{author}{\bibfnamefont{H.}~\bibnamefont{Pagels}} \bibnamefont{and}
  \bibinfo{author}{\bibfnamefont{S.}~\bibnamefont{Stokar}},
  \bibinfo{journal}{Phys. Rev. D} \textbf{\bibinfo{volume}{20}},
  \bibinfo{pages}{2947} (\bibinfo{year}{1979}).

\bibitem[{\citenamefont{Aoki et~al.}(1990)\citenamefont{Aoki, Bando, Kugo,
  Mitchard, and Nakatani}}]{Aoki:1990eq}
\bibinfo{author}{\bibfnamefont{K.-I.} \bibnamefont{Aoki}},
  \bibinfo{author}{\bibfnamefont{M.}~\bibnamefont{Bando}},
  \bibinfo{author}{\bibfnamefont{T.}~\bibnamefont{Kugo}},
  \bibinfo{author}{\bibfnamefont{M.~G.} \bibnamefont{Mitchard}},
  \bibnamefont{and} \bibinfo{author}{\bibfnamefont{H.}~\bibnamefont{Nakatani}},
  \bibinfo{journal}{Prog. Theor. Phys.} \textbf{\bibinfo{volume}{84}},
  \bibinfo{pages}{683} (\bibinfo{year}{1990}).

\bibitem[{\citenamefont{Holdom et~al.}(1990)\citenamefont{Holdom, Terning, and
  Verbeek}}]{Holdom:1990iq}
\bibinfo{author}{\bibfnamefont{B.}~\bibnamefont{Holdom}},
  \bibinfo{author}{\bibfnamefont{J.}~\bibnamefont{Terning}}, \bibnamefont{and}
  \bibinfo{author}{\bibfnamefont{K.}~\bibnamefont{Verbeek}},
  \bibinfo{journal}{Phys. Lett. B} \textbf{\bibinfo{volume}{245}},
  \bibinfo{pages}{612} (\bibinfo{year}{1990}).

\bibitem[{\citenamefont{Roberts and Williams}(1994)}]{Roberts:1994dr}
\bibinfo{author}{\bibfnamefont{C.~D.} \bibnamefont{Roberts}} \bibnamefont{and}
  \bibinfo{author}{\bibfnamefont{A.~G.} \bibnamefont{Williams}},
  \bibinfo{journal}{Prog. Part. Nucl. Phys.} \textbf{\bibinfo{volume}{33}},
  \bibinfo{pages}{477} (\bibinfo{year}{1994}), \eprint{hep-ph/9403224}.

\bibitem[{\citenamefont{Ma and Wang}(2003)}]{Ma:2003uv}
\bibinfo{author}{\bibfnamefont{Y.-L.} \bibnamefont{Ma}} \bibnamefont{and}
  \bibinfo{author}{\bibfnamefont{Q.}~\bibnamefont{Wang}},
  \bibinfo{journal}{Phys. Lett. B} \textbf{\bibinfo{volume}{560}},
  \bibinfo{pages}{188} (\bibinfo{year}{2003}), \eprint{hep-ph/0302143}.

\bibitem[{\citenamefont{Harada and Yamawaki}(1999)}]{Harada:1999zj}
\bibinfo{author}{\bibfnamefont{M.}~\bibnamefont{Harada}} \bibnamefont{and}
  \bibinfo{author}{\bibfnamefont{K.}~\bibnamefont{Yamawaki}},
  \bibinfo{journal}{Phys. Rev. Lett.} \textbf{\bibinfo{volume}{83}},
  \bibinfo{pages}{3374} (\bibinfo{year}{1999}), \eprint{hep-ph/9906445}.

\bibitem[{\citenamefont{Harada and
  Yamawaki}(2001{\natexlab{a}})}]{Harada:2000kb}
\bibinfo{author}{\bibfnamefont{M.}~\bibnamefont{Harada}} \bibnamefont{and}
  \bibinfo{author}{\bibfnamefont{K.}~\bibnamefont{Yamawaki}},
  \bibinfo{journal}{Phys. Rev. Lett.} \textbf{\bibinfo{volume}{86}},
  \bibinfo{pages}{757} (\bibinfo{year}{2001}{\natexlab{a}}),
  \eprint{hep-ph/0010207}.

\bibitem[{\citenamefont{Harada and
  Yamawaki}(2001{\natexlab{b}})}]{Harada:2001rf}
\bibinfo{author}{\bibfnamefont{M.}~\bibnamefont{Harada}} \bibnamefont{and}
  \bibinfo{author}{\bibfnamefont{K.}~\bibnamefont{Yamawaki}},
  \bibinfo{journal}{Phys. Rev. Lett.} \textbf{\bibinfo{volume}{87}},
  \bibinfo{pages}{152001} (\bibinfo{year}{2001}{\natexlab{b}}),
  \eprint{hep-ph/0105335}.

\bibitem[{\citenamefont{Harada and
  Yamawaki}(2001{\natexlab{c}})}]{Harada:2000at}
\bibinfo{author}{\bibfnamefont{M.}~\bibnamefont{Harada}} \bibnamefont{and}
  \bibinfo{author}{\bibfnamefont{K.}~\bibnamefont{Yamawaki}},
  \bibinfo{journal}{Phys. Rev. D} \textbf{\bibinfo{volume}{64}},
  \bibinfo{pages}{014023} (\bibinfo{year}{2001}{\natexlab{c}}),
  \eprint{hep-ph/0009163}.

\bibitem[{\citenamefont{Shifman et~al.}(1979)\citenamefont{Shifman, Vainshtein,
  and Zakharov}}]{Shifman:1978bx}
\bibinfo{author}{\bibfnamefont{M.~A.} \bibnamefont{Shifman}},
  \bibinfo{author}{\bibfnamefont{A.~I.} \bibnamefont{Vainshtein}},
  \bibnamefont{and} \bibinfo{author}{\bibfnamefont{V.~I.}
  \bibnamefont{Zakharov}}, \bibinfo{journal}{Nucl. Phys. B}
  \textbf{\bibinfo{volume}{147}}, \bibinfo{pages}{385} (\bibinfo{year}{1979}).

\bibitem[{\citenamefont{Workman et~al.}(2022)}]{ParticleDataGroup:2022pth}
\bibinfo{author}{\bibfnamefont{R.~L.} \bibnamefont{Workman}}
  \bibnamefont{et~al.} (\bibinfo{collaboration}{Particle Data Group}),
  \bibinfo{journal}{PTEP} \textbf{\bibinfo{volume}{2022}},
  \bibinfo{pages}{083C01} (\bibinfo{year}{2022}).

\bibitem[{\citenamefont{Pich}(2008)}]{Pich:2008xj}
\bibinfo{author}{\bibfnamefont{A.}~\bibnamefont{Pich}}, \bibinfo{journal}{PoS}
  \textbf{\bibinfo{volume}{CONFINEMENT8}}, \bibinfo{pages}{026}
  (\bibinfo{year}{2008}), \eprint{0812.2631}.

\end{thebibliography}

\end{document}